\documentclass[article,aps,prd]{revtex4}
\usepackage[colorlinks=true, pdfstartview=FitV, linkcolor=blue,
citecolor=red, urlcolor=magenta]{hyperref}
\usepackage[utf8]{inputenc}
\usepackage{graphicx}
\usepackage{latexsym}
\usepackage{amsmath}
\usepackage{amsfonts}
\usepackage{amssymb}
\usepackage{verbatim}
\usepackage{slashed}
\def\a{\alpha}

\def\b{\beta}
\def\m{\mu}
\def\s{\sigma}
\def\n{\nu}

\def\l{\lambda}
\def\lb{{\bar\lambda}}
\def\L{\Lambda}

\def\t{\theta}
\def\f{\phi}
\def\m{\mu}
\def\o{\omega}
\def\p{\pi}

\def\s{\sigma}

\def\x{\xi}

\def\L{\Lambda}
\def\O{\Omega}
\def\P{\Pi}

\def\S{\Sigma}

\def\X{\Xi}
\def\ovl{\overline}

\def\tb{\bar{\theta}}

\def\sb{\bar\sigma}

\def\d{\partial}

\def\f{\phi}

\def\P{\Psi}

\def\rd{\sqrt{2}}

\def\P{\Psi}
\def\p{\psi}

\def\dd #1 #2{{\delta #1\over \delta #2}}
\def\tr{{\rm tr}\ }
\def\til{\widetilde}

\def\ha{\frac{1}{2}}

\def\ov{\over}

\def\n{\nu}
\def\m{\mu}
\def\n{\nu}

\newcommand{\beq}{\begin{eqnarray}}
	\newcommand{\eeq}{\end{eqnarray}}

\begin{document}
	
	\title{RENORMALIZABILITY OF  $\mathcal{N}=1$ SUPER YANG-MILLS THEORY IN LANDAU GAUGE WITH A STUECKELBERG-LIKE FIELD
	}
	
	%%%%%%%%%%%%%%%%%%%%%%%%%%%%%%%%%%%%%%%%%%%%%%%%%%%%%%%%%%
	\author{$^{1}$M. A. L. Capri}
	\email{caprimarcio@gmail.com}
	\author{$^{1}$D.M. van Egmond}
	\email{duifjemaria@gmail.com}
	\author{$^{1}$M. S. Guimaraes}
	\email{msguimaraes@uerj.br}
	\author{$^{1}$O. Holanda}
	\email{ozorio.neto@uerj.br}
	\author{$^{1}$S. P.  Sorella}
	\email{silvio.sorella@gmail.com}
	\author{$^{1}$R. C. Terin}
	\email{rodrigoterin3003@gmail.com}
	\author{$^{1}$H. C. Toledo}
	\email{henriqcouto@gmail.com}
	\affiliation{$^{1}$Departamento de F\'isica Te\'orica, Rua S\~ao
		Francisco Xavier 524, Universidade do Estado do Rio de Janeiro,\\
		20550-013, Maracan\~a, Rio de Janeiro, Rio de Janeiro, Brazil.}

	%%%%%%%%%%%%%%%%%%%%%%%%%%%%%%%%%%%%%%%%%%%%%%%%%%%%%%%%%%
	
	%E-mail:{fabrito@df.ufcg.edu.br, leogrigorio@if.ufrj.br,
	%marceloguima@gmail.com, passos@fisica.ufpb.br, clovis@if.ufrj.br }
	%}
	
	%\preprint{}
	
	\begin{abstract}
		We construct a vector gauge invariant transverse field configuration $V^H$, consisting of the well-known superfield $V$ and of a Stueckelberg-like chiral superfield $\Xi$. The renormalizability of the Super Yang Mills action in the Landau gauge is analyzed in the presence of a gauge invariant mass term $m^2  \int dV  \mathcal{M}(V^H)$, with $\mathcal{M}(V^H)$ a power series in $V^H$. Unlike the original Stueckelberg action, the resulting action turns out to be renormalizable to all orders. 
	\end{abstract}
	\pacs{XX.XX, YY.YY} \maketitle
	
	%\vspace{1cm}

	\section{Introduction} 
	
	In this work we study the renormalizability properties of a ${\cal N} =1$ non-abelian gauge theory defined by a multiplet containing a massive vectorial excitation. The model we study is the supersymmetric version of a Stueckelberg-like action, in the sense that the massive gauge field is constructed by means of a compensating scalar field, thus preserving gauge invariance.   
	
	The history of Stueckelberg-like models is very well reviewed in \cite{Ruegg:2003ps}. Traditionally, most of the investigations have studied such models as potential alternative to the Higgs mechanism of mass generation, but as discussed in \cite{Delbourgo:1987np} there seems to be an unavoidable clash between renormalizability and unitarity in non-abelian Stueckelberg-like models. The original Stueckelberg model is abelian and has been rigorously proved \cite{Lowenstein:1972pr} to be renormalizable and unitary, but its non-abelian version is known to be perturbatively non-renormalizable \cite{Kunimasa:1967zza, Slavnov:1970tk, Slavnov:1972qb}. Physically, the problem is due to the high energy behavior of the longitudinal vector degree of freedom. In the abelian case it is perfectly compensated by the dynamics of the Stueckelberg field but in non-abelian theories this seems to be not so, resulting in incurable divergent interacting amplitudes or unbounded cross sections.

	Nevertheless there have been recent interests in the study of massive vector models without the Higgs. The main motivation comes here from the continuous efforts to understand the low energy behavior of strongly interacting gauge theories, such as QCD. Confinement is a  very important phenomenon in this context, but the physical mechanism behind it is still an open problem. A way to obtain information about this phenomenon is through lattice investigations which have revealed that the gluon propagator shows a massive behavior in the deep infrared non-perturbative region, while also displaying positivity violations which precludes a proper particle propagation interpretation \cite{Cucchieri:2004mf,  Cucchieri:2007md, Cucchieri:2007rg, Cucchieri:2008fc, Cucchieri:2008mv, Dudal:2013yva, Cornwall:2013zra}. Therefore, in a confining theory, the issue of the physical unitarity is a quite  complex and difficult topic. Of course, physical unitarity must hold in terms of the physical excitations of the spectrum which are bound states of quarks and gluons like, for instance,  mesons, barions, glueballs, etc. Though, the positivity violation of the two-point gluon correlation functions is taken as a strong  evidence of confinement, signalling that gluons are not excitations of the physical spectrum of the theory.  Nevertheless, renormalizability should be expected to hold since one wants to recover the good UV behavior of QCD. This trend of investigations led to many works that proposed modifications of the Yang-Mills theory to accommodate the lattice results \cite{Alkofer:2003jj, Dudal:2007cw, Dudal:2008sp, Tissier:2010ts, Strauss:2012dg}. Recent developments along these lines involve the introduction of modified Stueckelberg-like models \cite{Capri:2015ixa, Capri:2015nzw, Capri:2016aqq, Fiorentini:2016rwx, Capri:2016gut}  constructed as a generalization of a class of confining effective theories known as Gribov-Zwanziger scenarios \cite{Gribov:1977wm, Zwanziger:1988jt}, see \cite{Vandersickel:2012tz} for a review. Unlike the standard Stueckelberg action, these modified models enjoy the pleasant property of being renormalizable to all orders, see  \cite{Capri:2015ixa,  Capri:2015nzw, Capri:2016aqq, Fiorentini:2016rwx, Capri:2016gut} for a detailed account on the construction of these modified models and on their differences with the standard Stueckelberg theory. Let us also also mention here that, recently, a BRST invariant reformulation of the Gribov-Zwanziger theory has been achieved \cite{Capri:2016aqq, Capri:2016gut}, allowing its extension from the Landau gauge to an arbitrary covariant gauge.
	
 The present model is intended only to construct a renormalizable theory which generalizes the non-supersymmetric construction given in \cite{Fiorentini:2016rwx}. 
Issues like the perturbative unitarity of the models so obtained are not explicitly addressed. As far as we know, the non-supersymmetric model is not perturbative unitary. Though, it can be successfully  employed as an effective renormalizable model in order to investigate the non-perturbative infrared region of confining 
Yang-Mills theories. So far, the prediction of the non-supersymmetric model are in good agreement with the actual lattice data on the correlation functions of the theory, like the two-point gluon propagator. 

Our aim here is to construct a sypersimmetric generalization of this model for a future 
investigation of the confinement aspects of pure STM, which is known to be a 
confining theory. This is the main purpose of the present model. 

In a confining YM theory, the issue of the unitarity has to be faced through the
study of suitable colorless bound-state, a topic which is still too far from the goal  
of the present work, whose aim is that of obtaining a renormalizable massive 
SPYM theory whihc generalizes the model of \cite{Fiorentini:2016rwx}. 

	In this work we will carry out a supersymmetric generalization of the Stueckelberg-like model proposed in \cite{Capri:2015ixa, Capri:2015nzw, Capri:2016aqq, Fiorentini:2016rwx, Capri:2016gut}. We prove that the present supersymmetric generalization is renormalizable, a task that will be done by means of a set of suitable Ward identities. Supersymmetric generalizations of Stueckelberg-like models was studied since very early \cite{Delbourgo:1975uf} but mostly concentrated on the better behaved abelian models (see \cite{Kors:2004dx, Kors:2004ri, Kors:2005uz}, for instance, for a proposal of an abelian Stueckelberg sector in MSSM), with some constructions of non-abelian theories with tensor multiplets \cite{Nishino:2013kxa, Nishino:2013oea, Sezgin:2012ka} and also with composite gauge fields \cite{Nishino:2016xsw}.
	
	The work is organized as follows. In Section II we construct the $\mathcal{N}=1$ Supersymmetric massive classical action. In Section III we discuss the gauge fixing and the ensuing BRST symmetry. Sections IV and V are devoted to the derivation of a set of suitable Ward identities and to the characterization of the most general invariant local counterterm following the setup of the algebraic renormalization. In Section VI we provide a detailed analysis of the counterterm by showing that it can be reabsorbed into the starting classical action through a redefinition of the fields and parameters, thus establishing the all orders renormalizability of the model. Section VII contains our conclusion. The final Appendices collect the conventions and a few additional technical details.

	\section{ PURE  $\mathcal{N}=1$ SUSY STUECKELBERG-LIKE YANG-MILLS THEORY}
	
	In order to define the $\mathcal{N}=1$ Supersymmetric Stueckelberg-like Yang-Mills theory, we start with a real abelian gauge superfield, 
	\beq
	V(x,\t,\ovl\t)&=& C+i\t\chi-i\ovl\t\ovl\chi +\t\s^\m\ovl\t A_\m + {i\over 2} \t\t (M+iN) -{i\over 2} \ovl\t\ovl\t (M-iN)+ i \t\t\, \tb\left( \lb +{i\over 2}\sb^\m\d_\m\chi\right)\nonumber\\
	&-& i \ovl\t\ovl\t\, \t\left( \l -{i\over 2}\s^\m\d_\m\ovl\chi\right) +\ha \t\t\ovl\t\ovl\t \left( D-\ha \d^2 C\right),
	\label{V1}
	\eeq
	and with a  massless chiral superfield that acts as a Stueckelberg field
	\beq
	\X(x,\t,\ovl \t)&=&\x(x) + \rd\t\p(x) + i\t\s^\m\tb \d_\m \x(x) - \t\t f(x) -{i\over\rd} \t\t\d_\m\p(x) \s^\m\tb - {1\over 4}\t\t\ovl\t\ovl\t \d^2 \x(x)\\
	\ovl\X(x,\t,\ovl \t)&=&\ovl \x(x) + \rd\ovl\t\ovl\p(x) - i\t\s^\m\ovl\t \d_\m \ovl \x(x) - \ovl\t\ovl\t \,\ovl f(x) +{i\over\rd} \ovl\t\ovl\t \t\s^\m\d_\m\ovl\p(x) - {1\over 4}\t\t\ovl\t\ovl\t \d^2 \ovl \x(x).
	\eeq 
	It is then possible to construct a gauge-invariant superfield
	\beq
	V^H(x,\t,\ovl\t)&=&V(x,\t,\ovl\t)+i\X(x,\t,\ovl\t)-i\ovl\X(x,\t,\ovl\t),\label{eq4}
	\eeq
	which is invariant under the abelian gauge transformations
	
	\beq
	V\rightarrow V+i\f- i\ovl\f , \,\, \Xi \rightarrow \Xi - \phi, \,\, \ovl{\Xi} \rightarrow \ovl{\Xi} - \ovl{\phi}.\label{eq5}
	\eeq
	We now need a generalization of the definition of $V^H$ to the non-abelian case. We start with the gauge-invariant superfield \eqref{eq4}
	with every component now in the adjoint
	representation of the gauge group G, $V^H\equiv V^{Ha} T^a$, $(a=1,..., \mbox{dim}\, G)$ where the $T^a$   are the generators in the adjoint. Now, the fundamental object is $e^{V^H}$ instead of $V^H$. The non-abelian generalization of \eqref{eq4} is
	\beq\label{eq8}
	e^{V^H} = \overline{H} e^{V} H \ , \
	\eeq
	where $H=e^{i\X}$, $U=e^{i\f}$ and $V$ is a usual gauge superfield.
	The gauge transformations then have to be
	\beq
	e^{V}\to \overline{U}e^{V} U , \,\, H\to U^{-1}H, \,\, \overline{H}
	\to \overline{H} \ \overline{U}^{-1},\label{eq7}
	\eeq
	such that $V^H$ is gauge invariant. For infinitesimal transformations, this explicitly yields
	\beq
	\delta_{gauge} V&=&{i\ov 2}L_V(\f+\ovl\f) +{i\ov 2}(L_V\coth(L_{V/2}))(\f-\ovl\f)\nonumber\\
	&=&i(\f-\ovl\f)+{i\ov 2}[ V , \f+\ovl\f]+{i\ov 12}[V ,[V ,\f-\ovl\f]]+\mathcal{O}(V^3),\\
	\delta_{gauge}\Xi &=&{i\ov 2}L_\X \f -{1\ov 2}(L_\X\cot(L_{\X/2}))\f\nonumber\\
	&=&-\f +{i\ov 2}[ \X , \f]+{1\ov 12}[\X ,[\X ,\f]]+\mathcal{O}(\X^3),\\
	\delta_{gauge}\ovl\Xi &=&-{i\ov 2}L_{\ovl\X} \ovl\f -{1\ov 2}(L_{\ovl\X}\cot(L_{\ovl\X/2}))\ovl\f\nonumber\\
	&=&-\ovl\f -{i\ov 2}[ \ovl\Xi , \ovl\f]+{1\ov 12}[\ovl\X ,[\ovl\X ,\ovl\f]]+\mathcal{O}(\ovl\X^3),
	\eeq
	with $L_A X = [A,X]$. To first order (abelian gauge limit) in $\f$ this reproduces  \eqref{eq5}.

	Now, using all of the above definitions we can construct a gauge invariant $\mathcal{N}=1$ Supersymmetric Stueckelberg-like Yang-Mills model
	\beq\label{4gaugelagr}
	{\cal L}_{\rm SYM} &=&-{1\over 128 g^2}\tr\int dS \, W^{\a} W_{\a}\,+m^2 \, \tr \int dV \,{\cal{M}}(V^H),  
	\eeq
	with
	\beq
	W_{\a}\equiv \overline{D}^2(e^{-V}D_{\alpha}e^{V}) \;, \label{sspp}
	\eeq
	and
	\begin{eqnarray}
		\cal M&=& V^{Ha} V^{Ha} + \sigma_{1}^{abc} V^{Ha} V^{Hb} V^{Hc} +  \sigma_2^{abcd} V^{Ha} V^{Hb} V^{Hc} V^{Hd} +...\, , \label{tit}
	\end{eqnarray}
	where $\sigma_{1}^{abc}, \sigma_2^{abcd}, ...$ are a set of infinite arbitrary dimensionless parameters. As one can figure out, the fact that the generalized mass term ${\cal{M}}(V^H)$ is an infinite power series $V^H$ follows from the dimensionless character of $V^H$ itself. Though, from the pertubative point of view, only the first quadratic terms of the series \eqref{tit}, {\it i.e.} $m^2 V^{Ha} V^{Ha}$ will enter the expression of the superfield propagator. The remaining terms represent an infinite set of interaction vertices, a feature  which is typical of the non-abelian Stueckelberg-like theories.  
	
	Notice that the first term of the action, the pure supersymmetric Yang-Mills term $W^{\a} W_{\a}$, is invariant under $V \rightarrow V^H$. For more details about the $\mathcal{N}=1$ supersymmetry and conventions, see \cite{Piguet:1996ys,Grisaru:1983}.
	
	\section{Supersymmetric gauge invariant Stueckelberg-like Yang-Mills action in the Landau gauge}
	The supersymmetric extension of the Landau gauge is [1]
	\beq
	\ovl{D}^2D^2 V=0.
	\eeq
	We thus need to add the following terms to the action
	\beq
	{\cal L}_{SGF} = \frac{1}{8}Tr \int dS  A\ovl{D}^2D^2 V + c.c. = \frac{1}{8}Tr
	\int dV (AD^2 V +\ovl{A}\,\ovl{D}^2 V),
	\eeq
	where we introduced the auxiliary chiral superfield $A$, with
	the following field equations 
	\beq
	\frac{\delta }{\delta A}{\cal L}_{SGF}=\frac{1}{8}\ovl{D}^2D^2 V=0, \quad
	\frac{\delta }{\delta \ovl{A}}{\cal L}_{SGF}=\frac{1}{8}D^2\ovl{D}^2 V=0.
	\eeq
	Following the standard BRST procedure, the gauge fixing condition can be implemented in a BRST invariant way by defining the auxiliary field as the BRST variation of the anti-ghost field $c'$,, namely  $s c'=A$, so  that we can
	add the following BRST invariant term to the action in order to fix
	the gauge.
	\beq
	{\cal L}_{SGF} &=&
	s\left[\frac{1}{8}\tr\,\int dV (c' D^2
	V+c.c.\right]\nonumber\\
	&& = {1\over 8}\tr\int dS [(A \,\ovl D^2 D^2 V -c' \ovl D^2 D^2 sV+c.c.]\,.
	\label{kk}
	\eeq
	Looking at the gauge fixing \eqref{kk}, it is important to realize that for any local quantum field theory  involving dimensionless fields, one has the freedom of performing arbitrary re-parametrization of these fields. Examples  of this are the two-dimensional non-linear sigma model \cite{Blasi:1989},\cite{Becchi:1989} and quantum field theories with a Stueckelberg field entering the gauge fixing term \cite{Capri:2018}.  In the case of the gauge fixing \eqref{kk}, this means that we have the freedom of replacing $V^a$ by an arbitrary dimensionless function of $V^a$
	
	\beq
	V^a\rightarrow \mathcal{F}^a(V)=V^a + \a_1^{abc}V^b V^c +\a_2^{abcd}V^b V^c V^d + \a_3^{abcde}V^b V^c V^d V^e + ...\, , \;
	\label{eqq}
	\eeq
 where $\a_1^{abc}, \a_2^{abcd}, \a_3^{abcde}, ...$ are free dimensionless coefficients. This freedom, inherent to the dimensionless nature of $V^a$, is evident at the quantum
	level because of the fact that this field renormalizes in a non-linear way  \cite{Grisaru:1983,Piguet:1995er,Piguet1982,Piguet1984}. Therefore, \eqref{eqq} is expressing precisely the freedom one has in the choice of a re-parametrization for $V^a$. \\
	In our case, this means that instead of equation \eqref{kk} we could have just as well started with a term 
	
	\beq
	s\left[\frac{1}{8}\tr\,\int dV (c' D^2 V
	+c.c.)\right] \rightarrow s\left[\frac{1}{8}\tr\,\int dV (c' D^2
	\mathcal{F}(V)+c.c.)\right], 
	\label{tt}
	\eeq
	and this would not have affected the correlation functions of the gauge invariant quantities. The coefficients $(\a_1^{abc},\a_2^{abcd},\a_3^{abcde}..)$ are gauge parameters, not affecting the correlation functions of the gauge invariant quantities. The freedom that we have in the gauge fixing term will become apparent when performing the renormalization analysis. In fact, in section \ref{het}, we will use a generalized gauge-fixing term
	
	\beq
	S_{gf}^{gen}&=&s\left[\frac{1}{8}\tr\,\int dV (c' D^2
	\mathcal{F}(V)+c.c.\right] \nonumber \\
	&=& {1\over 8}\tr\int dS \left[A \,\ovl D^2 D^2 \mathcal{F}(V) -c' \ovl D^2 D^2 \frac{\partial \mathcal{F}(V)}{\partial V} sV\right] +c.c.\,,  \label{gfxgfx}
	\eeq 
	and by employing the corresponding  Ward identities, we can handle the ambiguity that is inherent to the gauge fixing. 
	The counterterm will then correspond to a renormalization of the gauge parameters $(\a_1^{abc},\a_2^{abcd},\a_3^{abcde}..)$, as will become clear in section \ref{is}.  \\

One of the striking features ensuring the renormalizability of the non-supersymmetric modified Stueckelberg-like models introduced in  \cite{Capri:2015ixa,  Capri:2015nzw, Capri:2016aqq, Fiorentini:2016rwx, Capri:2016gut,Capri:2018} was the implementation of a transversality constraint on the analogue of the gauge invariant field $V^H$. This transversality constraint gives rise to a deep difference between the modified models constructed in  \cite{Capri:2015ixa,  Capri:2015nzw, Capri:2016aqq, Fiorentini:2016rwx, Capri:2016gut,Capri:2018} and the conventional non-renormalizable Stueckelberg model. It is precisely the implementation of this transversality constraint which ensures the UV renormalizability of the modified model. We remind here the reader to reference \cite{Capri:2018} for a detailed account on the differences between the conventional and the modified Stueckelberg action. We then pursue here the same route outlined in \cite{Capri:2015ixa,  Capri:2015nzw, Capri:2016aqq, Fiorentini:2016rwx, Capri:2016gut,Capri:2018} and impose the transversality constraint also in the supersymmetric case. More precisely, this amounts to require that the superfield $V^H$ obeys the constraint

	\beq
	\ovl{D}^2D^2 V^H=0,   \label{tvh}
	\eeq
	which, at the level of the action, can be implemented by introducing the following terms 
	
	\beq
	{\cal L}_{T}=\frac{1}{8}\int dS\left( B\ovl{D}^2D^2 V^H -\eta'^a \ovl D^2 D^2 G^{a}_{V^H}(\eta,\ovl \eta)  \right) +c.c., 
	\eeq
	with $G^{a}_{V^H}(\eta,\ovl \eta)=(\eta^{a}-\ovl{\eta}^{a})+\frac{i}{2}f^{abc}V^{H,b}(\eta^{c}+\ovl \eta^{c})-\frac{i}{12}f^{amr}f^{mpq}(\eta^{p}-\ovl \eta^{p})V^{Hq} V^{Hr}+\mathcal{O}(V^{H3})$.
	 The field $B$ is a Lagrange multiplier implementing the transversality constraint \eqref{tvh}, while the fields $\eta,\ovl \eta$ are a set of ghost fields needed to compensate the Jacobian which arises from the functional integral over $B$ and $\Xi$ in order to get a unity, see \cite{Capri:2015ixa,  Capri:2015nzw, Capri:2016aqq, Fiorentini:2016rwx, Capri:2016gut,Capri:2018}  for the non-supersymmetric case. 
	\\
	
	Thus, adopting the Landau gauge, as well as the transversality condition, the total action becomes  
	
	\begin{eqnarray}
		{\Sigma}_{SPYM} & = & \int d^{4}x({\cal L}_{{\rm SYM}}+{\cal L}_{{\rm SP}}+{\cal L}_{SGF}+{\cal L}_{T})\nonumber \\
		& & =  -\frac{1}{128g^{2}}tr\int dS\, W^{a}W_{a}+{m^{2}\ov 2}\, tr\int dV\,{\cal M}(V^H) \nonumber \\
		&  & + \left[ s\left(\frac{1}{8}tr\,\int dS (c'\ovl D^2 D^{2}V )\right)+\frac{1}{8}\int dS \,\Bigg(B\ovl{D}^2D^2 V^H -\eta'^a \ovl D^2 D^2 G^{a}_{V^H}(\eta, \ovl \eta)  \Bigg) +c.c.\right]  
		 \nonumber \\
		& & =  -\frac{1}{128g^{2}}tr\int dS\, W^{a}W_{a}+{m^{2}\ov 2}\, tr\int dV\,{\cal M}(V^H) \nonumber \\
		& &
		+\left[ \frac{1}{8}\,\int dS\Bigg(A^{a}\ovl D^{2} D^{2}V^{a} -c'^{a}\ovl D^{2} D^{2}G_{V}^{a}\Bigg) +
		\frac{1}{8}\int dS \,\Bigg(B\ovl{D}^2D^2 V^H -\eta'^a \ovl D^2 D^2 G^{a}_{V^H}(\eta,\ovl \eta) \Bigg) +c.c. \right] 
		. \nonumber \\
		\label{o1}
	\end{eqnarray}
	
	This action enjoys the exact BRST nilpotent symmetry
	\begin{eqnarray}
		sV^{a} & = & G^{a}_{V}(c,\ovl c),\nonumber \\
		sV^{Ha} & = & 0, \nonumber \\
		s\Xi^{a} & = & G^{a}_{\Xi}(c),  \nonumber \\
		s\overline{\Xi}^{a} &  = & G^{a}_{\ovl \Xi}(\ovl c), \nonumber \\
		sc^{a} & =&-\frac{i}{2}f^{abc}c^{b}c^{c}, \nonumber \\
		s\eta^{a} & = & 0 \nonumber \\
		s\overline{c}^{a} & = &-\frac{i}{2}f^{abc}\ovl{c}^{b}\ovl{c}^{c} \nonumber \\
		s\overline{\eta}^{a} & = & 0 \nonumber \\
		sc'^{a} & = & A^{a}, \nonumber \\
		s A^a &=& 0,\nonumber \\
		s\eta'^{a} & = & 0, \nonumber \\
		s B^a &=&0, \nonumber \\
		s\ovl{c}'^{a} & = & \ovl{A}^{a}, \nonumber \\
		s\ovl{\eta}'^{a} & = & 0, \nonumber \\
		\mbox{with}\nonumber  \\
		G^{a}_{V}(c,\ovl c)&=&(c^{a}-\ovl{c}^{a})-\frac{1}{2}f^{abc}V^{b}(c^{c}+\ovl c^{c})-\frac{i}{12}f^{amr}f^{mpq}(c^{p}-\ovl c^{p})V^q V^r+\mathcal{O}(V^3), \nonumber \\
		G^{a}_{\Xi}(c)&=&-c^{a}-\frac{1}{2}f^{abc}\Xi^{b}c^{c}-\frac{1}{12} f^{amr} f^{mpq} c^p \Xi^q \Xi^r+\mathcal{O}(\Xi^3), \nonumber \\
    		G^{a}_{\ovl \Xi}(\ovl c)&=& -\ovl c^{a}+\frac{1}{2}f^{abc}\ovl\Xi^{b}\ovl c^{c} -\frac{1}{12} f^{amr} f^{mpq} c^p \ovl \Xi^q \ovl \Xi^r + \mathcal{O}(\ovl \Xi^3) 
		  \label{brstbrst}
	\end{eqnarray}
	
	and  
	\begin{equation}
		s{\mathcal{S}}_{SPYM}=0\,, \qquad s^2=0  \;. 
	\end{equation}
	
	\section{Renormalizability analysis}
	
	In order to analyze the renormalizability of the action \eqref{o1}, we start by establishing the set
	of Ward identities that will be employed for the study of the quantum corrections. Following the algebraic renormalization procedure \cite{Piguet:1995er}, we  have first to add some external sources coupling to non-linear BRST transformations of the fields and of the composite operators entering the classical action.  Therefore, we need to introduce a set of external BRST invariant sources $(\O^a, \L^a,\ovl \L^a, L^a,\ovl L^a )$ coupled to the non-linear BRST variations of  $(V^a,\X^a, \ovl \X^a, c^a, \ovl c^a)$ as well as sources $(\Pi^a, \Psi^a)$ coupled to the BRST invariant composite operators $(V^H, G^a_{V^H})$,

	\begin{eqnarray}
		s\Omega^{a}  = s\P^a = s\Pi^a = s\Lambda^{a}=s\ovl{\Lambda}^{a}=s L^{a} = s\ovl{L}^{a}=0 .
	\end{eqnarray}
	
	We shall thus start with the BRST invariant complete action $\S$ defined by
	
	\begin{eqnarray}
		\Sigma_{SPYM} & = & \Sigma_{{\rm SYM}}+\Sigma_{{\rm SP}}+\Sigma_{SGF}+\Sigma_{T}+\Sigma_{EXT}
	\end{eqnarray}
	\begin{eqnarray}
		\Sigma_{{\rm SPYM}} & = & -\frac{1}{128g^{2}}tr\int dS\,W^{a}W_{a}+\frac{m^2}{2}\int dV \cal{M}(V^H) \nonumber \\
		&  & +\frac{1}{8}\,\int dS\Bigg\{ A^{a}\ovl D^2 D^{2}V^{a}-c'^{a}\ovl D^2 D^{2}G_{V}^a \Bigg\} +c.c.\nonumber \\
		&&+\frac{1}{8}\int dS \,\Bigg\{B\ovl{D}^2D^2 V^H -\eta'^a\ovl D^2 D^2 G^{a}_{V^H}  
		\Bigg\} +c.c. \nonumber \\
		&  &+ \int dV\Bigg\{\Omega^{a}G_{V}^a +\Pi^a V^{H,a}+\P^{a}G_{V^H}^a\Bigg\} \nonumber \\
		&  & +\int dS\Bigg\{-\Lambda^{a}G_{\Xi}^a  -\frac{i}{2}f^{acb}L^{a}c^{b}c^{c}\Bigg\} +c.c. .\label{fullaction}
	\end{eqnarray}
	%where in the physical limit the sources are
	
	%\begin{eqnarray}
		%\O^a=\Pi^a=\Psi^a=\Lambda^a=L^a=0\\
	%\end{eqnarray}
	%to recover \eqref{o1}. 
	All quantum numbers, dimensions and $R$-weights of all fields and sources are displayed in tables \ref{tb1}
	and \ref{tb2}.
	
	\begin{table}[!h]
		\centering
		\caption{Quantum numbers of the fields}
		\label{my-label}
		\begin{tabular}{|l|l|l|l|l|l|l|l|l|l|l|l|l|l|l|l|l|l|l|l|}
			\hline
			& $\theta ^a$ & $D^a$ & $V^a$ & $V^{Ha}$ & $\Xi^a$  & $\overline{\Xi}^a$ & $c^a$ & $\bar{c}^a$ & $c'^a$ & $\bar{c}'^a$ & $A^a$ & $B^a$ & $\eta^a$ & $\eta'^a$ & $\bar{\eta}^a$ & $\bar{\eta}'^a $ \\ \hline
			dimension & -1/2 & 1/2 &0& 0 & 0 & 0 & 0 & 0 & 1 & 1 & 1 & 1 & 0 & 1& 0 & 1\\ \hline
			c-ghost \# &-1& 0 & 0 & 0 & 0 & 0 & 1 & 1 & -1 & -1 & 0 & 0 & 0 & 0 & 0 & 0    \\ \hline
			$\eta$ -ghost \# &-1& 0 & 0 & 0 & 0 & 0 & 0 & 0 & 0 & 0 & 0 & 0 & 1 & -1 & 1 & -1    \\ \hline
			R-weight &-1 & 1& 0 & 0 & 0 & 0 & 0 & 0 & -2 & 2 & -2 & -2 & 0 & -2 & 0 & 2 \\ \hline
		\end{tabular}
		\label{tb1}
	\end{table}

	\begin{table}[!h]
		\centering
		\caption{Quantum numbers of the sources}
		\label{my-label}
		\begin{tabular}{|l|l|l|l|l|l|l|l|l|l|l|l|l|l|l|l|l|l|l|l|}
			\hline
			& $\Omega^a$ & $\Psi^a$ & $\ovl{\Lambda}^a$ & $\Lambda^a$ & $L^a$ & $\ovl{L}^a$   & $\Pi^a$ \\ \hline
			dimension & 2 & 2 & 3 & 3 & 3  & 3  & 2\\ \hline
			c-ghost \#& -1 & 0 & -1 & -1 & -2   & -2   & 0 \\ \hline
			$\eta$-ghost \#& 0 & -1 & -1 & -1 & 0  & 0  & 0 \\ \hline
			R-weight & 0 & 0 & 2 & -2 & -2  & 2  & 0  \\ \hline
		\end{tabular}
		\label{tb2}
	\end{table}

	\subsection{Ward identities and algebraic characterization of the invariant counterterm}

	The complete action $\Sigma_{SPYM}$ obeys a large set of Ward identities, being:  
	
	\begin{itemize}
		\item The Slavnov-Taylor identity: 
		
		\begin{eqnarray}
			\mathcal{S}(\Sigma) & = & \int dV\frac{\delta\Sigma}{\delta\Omega^{a}}\frac{\delta\Sigma}{\delta V^{a}} +\Bigg[ \int dS \left(\frac{\delta\Sigma}{\delta\Lambda^{a}}\frac{\delta\Sigma}{\delta\Xi^{a}}+\frac{\delta\Sigma}{\delta L^{a}}\frac{\delta\Sigma}{\delta c^{a}}+A^{a}\frac{\delta\Sigma}{\delta c'^{a}}\right) +c.c.\Bigg]= 0
			\label{bgn}
		\end{eqnarray}
		
		\item The gauge-fixing equations:
		
		\beq
		\frac{\delta \Sigma}{\delta A^a} &=& \frac{1}{8} \bar{D}^2{D}^2 V^a, \qquad \frac{\delta \Sigma}{\delta \bar{A}}= \frac{1}{8} D^2 \bar{D}^2 V^a 
		\label{gff}
		\eeq
		
		\item The equation for the Lagrange multiplier $B^a$:
		\beq
		\frac{\delta \Sigma}{\delta B^a} &=& \frac{1}{8} \bar{D}^2{D}^2 \frac{\delta \Sigma}{\delta\Pi^a}, \qquad \frac{\delta \Sigma}{\delta \bar{B}^{a}}= \frac{1}{8} D^2 \bar{D}^2 \frac{\delta \Sigma}{\delta\Pi^a}
		\label{w2}
		\eeq
		
		\item The anti-ghost equations:
		
		\begin{align}
			\mathcal{G}_-^a \Sigma =0, \qquad \bar{\mathcal{G}}_-^a \Sigma =0 ,
			\label{ag}
		\end{align}
		with

		\begin{eqnarray}
			\mathcal{G}_-^a= \frac{\delta}{\delta c'^a}+\frac{1}{8} \bar{D}^2 D^2 \frac{\delta }{\delta \Omega^a} \quad \text{and} \quad \bar{\mathcal{G}}_-^a= \frac{\delta}{\delta \bar{c}'^a}+\frac{1}{8} D^2 \bar{D}^2  \frac{\delta }{\delta \Omega^a} 
		\end{eqnarray}
		
		\item The $\eta$ Ward identities
		
		\beq
		\mathcal{F}_-^a \Sigma =0, \qquad {\mathcal{\ovl F}}_{-}^a \Sigma =0 ,
		\eeq
		with

		\beq
		\mathcal{F}_-^a= \frac{\delta}{\delta \eta'^a}+\frac{1}{8} \bar{D}^2 D^2 \frac{\delta }{\delta \Psi^a} \quad \text{and} \quad {\mathcal{\ovl F}}_{-}^a= \frac{\delta}{\delta \ovl{\eta}'^a}+\frac{1}{8} D^2 \bar{D}^2  \frac{\delta }{\delta \Psi^a} 
		\label{w1}
		\eeq
		
		\item  The linearly broken ghost equation \cite{Piguet:1995zz}
		
		\begin{align}
			\mathcal{G}_+ \Sigma = \Delta_{clas},
			\label{gh}
		\end{align}
		
		with 
		
	\begin{eqnarray}
			\mathcal{G}_+ &&= \int dS \Bigg( \frac{\delta}{\delta c^c}-if^{abc}c'^{a}\frac{\delta}{\delta A^b}\Bigg) + \int d\ovl{S} \Bigg(\frac{\delta}{\delta \ovl{c}^c}-if^{abc}\ovl{c}'^{a}\frac{\delta}{\delta \ovl{A}^b} \Bigg),
		\end{eqnarray}
		
		and
		
		\begin{eqnarray}
			\Delta_{clas}=if^{abc}\int dV \Omega^a V^b+i f^{abc}\int dSL^ac^b +if^{abc}\int d\ovl{S}\,\ovl{L}^a \ovl{c}^b.
		\end{eqnarray}
	Notice that the breaking term 	$\Delta_{clas}$ is purely linear in the quantum fields. As such, it will be not affected by the quantum corrections  \cite{Piguet:1995zz,Piguet:1995er}.

		\item The linear symmetries under supersymmetry, translations, $R$-transformations and rigid transformations are expressed by the Ward identities
		
		\begin{align}
			W_X \Sigma =-i\sum_{\phi} \int \delta_X \phi \frac{\delta}{\delta\phi}\Sigma=0, \quad X=Q_{\alpha}, P_{\mu}, R \text{ and rigid transformations} 
			\label{end}
		\end{align}
		
	\end{itemize}
such that $\delta_{\mu}^P$, $\delta_{\a}^{Q}$, $\delta_{\dot\a}^{\ovl Q}$, $\delta^{R}$ are defined by appendix \ref{apb}. Thus, we can see that the covariant action $\S$ satisfy the Ward identities and Lorentz invariance. 
	\section{The algebraic characterization of the invariant counterterm and renormalizability \label{het}}
	
	In order to characterize the most general invariant counterterm which can be freely added to
	all order in perturbation theory, we follow the setup of the algebraic renormalization \cite{Piguet:1995er} and
	perturb the classical action \eqref{o1}, by adding an integrated local quantity in the fields
	and sources, $\S_{CT}$ , that has  $R$-weight 0, ghost number (0,0), is hermitian and has dimension 3 in case of a chiral superfield, or 2 in case of a vector superfield. We demand
	thus that the perturbed action, $(\S + \varepsilon \S_{CT})$, where $\varepsilon$ is an expansion parameter, fulfills, to
	the first order in $\varepsilon$, the same Ward identities obeyed by the classical action $\S$, i.e. equations \eqref{bgn} to \eqref{end}. This amounts to impose the following constraints on $\Sigma_{CT}$:
	
	\begin{eqnarray}
		\mathcal{B}_{\S}\S_{ct}&=&0, \label{pp}\\
		\frac{\delta \S_{ct}}{\delta A^a }&=&\frac{\delta \S_{ct}}{\delta \ovl A^a }=0 \label{ik}\\
		\frac{\delta \Sigma_{CT}}{\delta B^a} &=& \frac{1}{8} \ovl {D}^2{D}^2 \frac{\delta \Sigma}{\delta\Pi^a},
		\qquad \frac{\delta \Sigma_{CT}}{\delta \ovl {B}^{a}}=\frac{1}{8} D^2 \ovl {D}^2 \frac{\delta \Sigma}{\delta\Pi^a}\label{01}\\
		\mathcal{G}_{-}^a \S_{CT} &=& \ovl{ \mathcal{G}}_{-}^a \S_{CT}=0 \label{00}\\
		\mathcal{F}_{-}^a \S_{CT}&=&\ovl{\mathcal{F}}_{-}^a\S_{CT}=0\\
		\mathcal{G}_{+}^a \S_{CT}&=&0,
	\end{eqnarray}
	where $\mathcal{B}_{\S}$ is the so-called nilpotent linearized Slavnov-Taylor operator \cite{Piguet:1995er}, defined as 
	\begin{eqnarray}
		\mathcal{B}_{\Sigma} = \int dV\left\{ \frac{\delta\Sigma}{\delta\Omega^{a}}\frac{\delta}{\delta V^{a}}+\frac{\delta\Sigma}{\delta V^{a}}\frac{\delta}{\delta\Omega^{a}}\right\}  +\int dS\left\{ \frac{\delta\Sigma}{\delta\Lambda^{a}}\frac{\delta}{\delta\Xi^{a}}+\frac{\delta\Sigma}{\delta\Xi^{a}}\frac{\delta}{\delta\Lambda^{a}}+\frac{\delta\Sigma}{\delta L^{a}}\frac{\delta}{\delta c^{a}}+\frac{\delta\Sigma}{\delta c^{a}}\frac{\delta}{\delta L^{a}}+A^{a}\frac{\delta}{\delta c'^{a}}+c.c.\right\},
	\end{eqnarray}
	with $\mathcal{B}_{\S}\mathcal{B}_{\S}=0$. 
	From equation \eqref{pp} one learns that $\S_{CT}$ belongs to the cohomology \cite{Piguet:1995er} of the linearized Slavnov-
	Taylor operator $\mathcal{B}_{\S}$ in the space of the integrated local quantities in the fields and sources  with ghost number (0,0), $R$-weight 0 and dimension 3 in case of a chiral superfield, or 2 in case of a vector superfield. Therefore, we can set
	\beq
	\S_{CT}=\Delta_{cohom}+\mathcal{B}_{\S}\Delta^{-1},
	\label{ml}
	\eeq
	where $\Delta^{(-1)}$ denotes a zero-dimensional integrated quantity in the fields and sources with ghost number (-1,0) and $R$-weight 0. The term $\mathcal{B}_{\S}\Delta^{(-1)}$ in equation \eqref{ml} corresponds to the trivial solution, i.e. to the
	exact part of the cohomology of $\mathcal{B}_{\S}$. On the other hand, the quantity $\Delta_{cohom}$ identifies the non-trivial solution,
	i.e. the cohomology of $\mathcal{B}_{\S}$ , meaning that $\Delta_{cohom} \neq \mathcal{B}_{\S}Q$, for any local integrated Q.
	
	In its most general form, $\Delta_{cohom}$ is given by 
	\beq
	\Delta_{cohom}&=&\int dS{a_0 (V^H) \ov 128g^2} W^{\a}W_{\a}+\int dV\, {m^2 \ov 2}\,{\til{\cal{{M}}} }(V^H) \nonumber  \\
	&+&\bigg(\int dS\,b_1^{ab}(V^H)\, \,B^a\ovl{D}^2D^2 V^{H,b}+\int dS \,b_2^{ab}(V^H)\,V^{H,b}\ovl{D}^2D^2 B^a+\int dS \,\,B^a V^{H,b}\ovl{D}^2D^2 b_3^{ab}(V^H)  \nonumber \\
	&+&\int dS\,d_{1}^{ab}\,\eta'^a\ovl D^2 D^2 \eta^b + \int dS\, c_{1}^{abc}(V^H)\eta'^a  V^{H,b}\ovl D^2 D^2 \eta^c +\int dS\, c_{2}^{abc}(V^H)\eta'^a  \eta^b \ovl D^2 D^2 V^{H,c} \nonumber \\
	&+&\int dS\, c_{3}^{abc}(V^H) \eta^a V^{H,b} \ovl D^2 D^2 \eta'^{c}+\int dS\, \eta'^a\eta^b  V^{H,c}\ovl D^2 D^2 c_{4}^{abc}(V^H)+\int dS\, c_{5}^{abc}(V^H)\eta'^a  \ovl\eta^b \ovl D^2 D^2 V^{H,c} \nonumber \\
	&+&\int dS\, c_{6}^{abc}(V^H) \ovl\eta^a V^{H,b} \ovl D^2 D^2 \eta'^{c}+\int dS\, \eta'^a\ovl\eta^b  V^{H,c}\ovl D^2 D^2 c_{7}^{abc}(V^H)+c.c. \bigg) \nonumber \\
	&+&\int dV b_{4}^{ab}(V^H)\Pi^{a}V^{H,b}+\int dV d_2^{ab} \Psi^a \eta^b \nonumber \\
	&+& \int dV c_8^{abc}(V^H)V^{Hc}\P^{a}\eta^{b}+\int dV d_3^{ab} \Psi^a \ovl \eta^b +\int dV c_9^{abc}(V^H)V^c\P^{a}\ovl \eta^{b}, \label{cohcoh}
	\eeq
	with 
	\begin{eqnarray}
		\til{\cal M}&=& b_0 V^{Ha} V^{Ha} + \til{\sigma}_{1}^{abc} V^{Ha} V^{Hb} V^{Hc} +  \til{\sigma}_2^{abcd} V^{Ha} V^{Hb} V^{Hc} V^{Hd} +... \, ,
	\end{eqnarray}
	and $(a_0 (V^H),b_i^{ab}(V^H),c_i^{abc}(V^H),d_i^{ab})$ arbitrary coefficients. 
	Then, after implementing the constraints \eqref{01} and \eqref{00} we find
	\beq
	\Delta_{cohom}&=&\int dS{a_0 (V^H)\ov 128g^2} W^{\a}W_{\a}+\int dV\, {m^2 \ov 2}\,{\til{\cal{{M}}} }(V^H)  \nonumber \\
	&+& \int dV d_1^{ab}\Psi^a \eta^b+\left(\frac{1}{8}\int dS \,\,d_1^{ab}\eta^a \ovl D^2 D^2 \eta'^b+c.c.\right) \nonumber \\
	&+& \int dV d_2^{ab}\Psi^a \ovl \eta^b+\left(\frac{1}{8}\int dS \,\,d_2^{ab}\ovl \eta^a \ovl D^2 D^2 \eta'^b+c.c.\right) \nonumber \\
	&+&\int dV c_1^{abc}(V^H) \Psi^a V^{Hb}\eta^c+ \left(\frac{1}{8} \int dS c_1^{abc}(V^H)\,\,\eta^a V^{Hb}\ovl D^2 D^2 \eta'^c+c.c.\right) \nonumber \\
	&+&\int dV c_2^{abc}(V^H) \Psi^a V^{Hb}\ovl \eta^c+ \left(\frac{1}{8} \int dS c_2^{abc}(V^H)\,\, \ovl \eta^a V^{Hb}\ovl D^2 D^2 \eta'^c +c.c.\right) \nonumber \\
	&+&\int dV b_1^{ab}(V^{H}) V^{Hb}\Pi^a+\left(\frac{1}{8} \int dS \,\,b_1^{ab}(V^H)V^{Hb}\ovl D^2 D^2 B^a+c.c.\right).
	\label{lgl}
	\eeq
	Let us now discuss the trivial part of the counterterm, $\mathcal{B}_{\S}\Delta^{(-1)}$. The term $\Delta^{(-1)}$, taking into account the
	quantum numbers of the fields and sources, can be parametrized in its most general form as:
	\beq
	\Delta^{-1}&=& \int dV \Bigg( F_{1}^{ab}V^b \O^a + F_{2}^{ab}V^b D^2 c'^a +F_{3}^{ab}D^2 V^b c'^a+D^2 F_{4}^{ab} V^b c'^a + c.c.\Bigg) \nonumber\\
	&&+\left[\int dS \left( a_1^{ab} L^a c^b +a_{2}^{ab}\Xi^{b}\Lambda^{a}\right) +c.c.\right],
	\label{kn}
	\eeq
	where,
	\begin{eqnarray}
		F_{1,..,4}^{a} & = & F_{1,..,4}^{a}\left[V, \Xi,\ovl{\Xi}\right] \nonumber \\
		a_{1}^{ab} & = & a_{1}^{ab}\left[\Xi\right] \nonumber \\
		a_{2}^{ab} & = & a_{2}^{ab}\left[{\Xi}\right]. \label{yyyy}
	\end{eqnarray}
	Imposing the constraint \eqref{ik}
	\beq
	\frac{\delta}{\delta A^a }\mathcal{B}_{\S}\Delta^{(-1)}=\frac{\delta}{\delta \ovl A^a }\mathcal{B}_{\S}\Delta^{(-1)}=0,
	\label{iii}
	\eeq
	and observing from eq. \eqref{kn} that 
	\beq
	\frac{\delta \Delta^{(-1)}}{\delta A^a}=\frac{\delta \Delta^{(-1)}}{\delta \ovl A^a}=0 \,\,\, \Rightarrow \,\,\, \mathcal{B}_{\S} \frac{\delta \Delta^{(-1)}}{\delta A^a}=\mathcal{B}_{\S} \frac{\delta \Delta^{(-1)}}{\delta \ovl A^a}=0,
	\eeq
	we can use the relation
	\beq
	[\frac{\delta }{\delta A^a}, \mathcal{B}_{\Sigma} ]=\frac{\delta}{\delta c'^a}+\frac{1}{8}\ovl D^2 D^2 \frac{\delta}{\delta \O},
	\eeq
	to impose 
	\beq
	\left(\frac{\delta}{\delta c'^a}+\frac{1}{8}\ovl D^2 D^2 \frac{\delta}{\delta \O}\right)\Delta^{-1}&=& \ovl D^2 D^2(F_2^{ab} V^b)+\ovl D^2 (F_3^{ab} D^2 V^b)+\ovl D^2 \left((D^2 F_4^{ab})V^b\right)+\frac{1}{8}\ovl D^2 D^2 (F_1^{ab} V^b)=0. \label{mm}
	\eeq
	From eq. \eqref{mm} we find the relations 
	\beq
	F_2^{ab}&=&-\frac{1}{8}F_1^{ab} \nonumber \\
	F_3^{ab}&=&0 \nonumber \\
	F_4^{ab}&=&a \delta^{ab} \;, \label{f4444}
	\eeq  
	so that 
	\beq
	\Delta^{-1}&=& \int dV \Bigg(F_{1}^{ab}V^b(\O^a-\frac{1}{8} D^2 c'^a  -\frac{1}{8}\ovl D^2 \ovl c'^a)\Bigg) \nonumber \\
	&&+ \left[ \int dS \Bigg( a_1^{ab} L^a c^b + a_{2}^{ab}\Xi^{b}\Lambda^{a}\Bigg)+c.c.\right] .
	\label{do}
	\eeq
	\\
	\\
	We can further reduce the number of parameters in $\S_{CT}$ by noticing that if we set \cite{Fiorentini:2016rwx}
	\begin{eqnarray}
		m^2=\Pi=\Psi=\Lambda=\ovl \Lambda=0,
		\label{limit}
	\end{eqnarray}
	in \eqref{fullaction}, the resulting action is	
	\begin{eqnarray}
		\Sigma_{{\rm SPYM}} & = & -\frac{1}{128g^{2}}tr\int dS\,W^{a}W_{a} \nonumber \\
		&  & +\frac{1}{8}\,\left[\int dS\Bigg( A^{a}\ovl D^2 D^{2}V^{a}-c'^{a}\ovl D^2 D^{2}G_{V}^a \Bigg) +c.c.\right]\nonumber \\
		&&+\frac{1}{8}\left[ \int dS \,\Bigg( B^{a}\ovl{D}^2D^2 V^{H,a} -\eta'^a\ovl D^2 D^2 G^{a}_{V^H} \Bigg) +c.c.\right]\nonumber \\
		&&+ \int dV\,\Omega^{a}G_{V}^a +\left[ \int dS\Bigg(-\frac{i}{2}f^{acb}L^{a}c^{b}c^{c}\Bigg) +c.c.\right] ,  \label{spyappb}
	\end{eqnarray}
	which is nothing but the Super Yang-Mills gauge-fixed action in the Landau gauge (see appendix \ref{apc}), with the addition of the following terms
	\begin{eqnarray}
\Sigma_{B}=		\frac{1}{8}\left[\int dS \,\Bigg( B\ovl{D}^2D^2 V^H -\eta'^a\ovl D^2 D^2 G^{a}_{V^H} \Bigg) +c.c.\right] .
	\end{eqnarray}
However, upon integration over $(B,\eta, \ovl \eta,\Xi)$, these terms give rise to a unity. Thus, in the limit \eqref{limit} the starting action takes the following form:
\beq
\Sigma_{inicial} = \Sigma_{SPYM} + \Sigma_B .
\eeq
Let us consider now the correlation functions of the Yang-Mills superfield $V$, namely 
\beq
\langle V(1)........V(n)\rangle = \frac{\int [D\Phi]  \langle V(1)........V(n)\rangle  exp(- \Sigma_{inicial} )}{  \int [D\Phi]    exp(- \Sigma_{inicial} )}, \label{inicial}
\eeq
where $[D\Phi]$ stands for integration over all fields. Though, since $m^2=0$,  one can directly 
perform in \eqref{inicial} the integration over $(H, B, \eta, \ovl \eta)$, i.e. of the fields appearing in $\Sigma_{B}$.

This integration is easily seen to give a unity. It is in fact nothing but a  Super Faddeev-Popov term (see \cite{Grisaru:1983}) which, due to $m^2=0$, gives a unity. 

Therefore, in the limit, $m^2=0$, it follows that 
\beq
\langle V(1)........V(n)\rangle = \frac{\int [D\Phi]  <V(1)........V(n)>  exp(- \Sigma_{SPYM} )}{  \int [D\Phi]    exp(- \Sigma_{SPYM} ) },
\eeq
meaning that the correlator $\langle V(1)........V(n)\rangle$ reduces to that of standard SPYM. As consequence, the dimensionless, and thus 
$m$-independent,  coefficients appearing in the counterterm $\Sigma_{CT}$  are subject to the following additional conditions 
%, the counterterm $\S_{CT}$ has to reduce to the standard counterterm of $\mathcal{N}=1$ Super Yang-Mills, which we can be found in appendix \ref{apc}.   Implementing this condition, we find the following restrictions on the coefficients in equation \eqref{lgl}
	\begin{eqnarray}
		a_0(V^H)&=&a_0, \nonumber \\
		d_1^{ab}\eta^b+d_2^{ab}\ovl \eta^b+(c_1^{abc}\eta^c+c_2^{abc} \ovl \eta^c)V^{Hb}&=&b_1 G_{V^H}^a(\eta,\ovl \eta), \nonumber \\
		b_1^{ab}(V^H)&=&b_1 \delta^{ab} , \label{zzzz}
	\end{eqnarray}
	and for equation \eqref{kn}
	\begin{eqnarray}
		F_{1,..,4}^{a}\left[V, \Xi,\ovl{\Xi}\right] & = & F_{1,..,4}^{a}\left[V \right]\nonumber \\
		a_{1}^{ab} \left[\Xi\right]& = & a_{1}^{ab}\nonumber \\
		a_{2}^{ab} \left[\Xi\right] & = & a_{2}^{ab}\left[{\Xi}\right], \label{bbbb}
	\end{eqnarray}
	so that 	
	\begin{eqnarray}
		\Delta_{cohom}&=& \int dS \,\frac{a_0}{128g^2}W^{\a}W_{\a}+\int dV\, {m^2 \ov 2}\,{\til{\cal{{M}}} }(V^H) \nonumber \\
		&+&\int dV \,b_1 V^{Ha} \Pi^a+\frac{1}{8} \left[ \int dS \,b_1\left(V^{Ha} \ovl D^2 D^2B^a -\eta'^a \ovl D^2 D^2 G^a_{V^H}(\eta, \ovl \eta)\right) +c.c.\right] \nonumber \\
		&+&\int dV b_1\Psi^a G_{V^H}^a(\eta,\ovl \eta),
		\label{d}
	\end{eqnarray}
	and 
	\beq
	\Delta^{-1}&=& \int dV \Bigg(F_{1}^{ab}(V)V^b(\O^a-\frac{1}{8} D^2 c'^a  -\frac{1}{8}\ovl D^2 \ovl c'^a)\Bigg) \nonumber \\
	&&+\left[ \int dS \Bigg( a_1^{ab} L^a c^b + a_{2}^{ab}(\Xi)\Xi^{b}\Lambda^{a}\Bigg) +c.c. \right].
	\label{do}
	\eeq
	For the purpose of the analysis of the renormalization factors at the end of this section, we will rewrite the counterterm $\S_{CT}$ of the action in its parametrized form, namely as contact terms written in terms of the starting classical action $\Sigma$, being  given by the following expression
	\begin{eqnarray}
		\Sigma_{CT} &=&\int dS a_0 g^2 \frac{\delta \Sigma}{\d g^2} + \int dV b_0 m^2 \frac{\delta \S}{\delta m^2}+b_1 \left[ \int dS \left(B^a \frac{\d\S}{\d B^a}\right) +c.c.\right] + b_1 \int dV \, \Pi^a \frac{\d\S}{\d \Pi^a} \nonumber \\
		&+&\frac{1}{2}b_1 \int dV \Psi^a \frac{\d\S}{\d \Psi^a}+\frac{1}{2}b_1 \left[ \int dS \left(\eta^a\frac{\d\S}{\d \eta^a}\right) +c.c.\right]+\frac{1}{2}b_1 \left[ \int dS \left(\eta'^a\frac{\d\S}{\d \eta'^a}\right) +c.c.\right] \nonumber \\
		&+&\int dV \left(\tilde{\s}_1^{abc}\frac{\delta \Sigma}{\delta \sigma_1^{abc}}+\tilde{\s}_2^{abcd}\frac{\delta \Sigma}{\delta \sigma^{abcd}_2}+\tilde{\s}_3^{abcde}\frac{\delta \Sigma}{\delta \sigma^{abcde}_3}+...\right) \nonumber \\
		&+& \int dV\Bigg(\left({\delta F_{1}^{ab}\ov\delta V^c}V^b+F_1^{ab}\delta^{bc}\right) {\O}^a{\delta\S\ov\delta\O^c}+F_{1}^{ab}V^b{\delta\S\ov\delta V^a} \Bigg) \nonumber \\
		&+&\left[\int dS \Bigg(a_2^{ab}\Xi^b{\delta \S\ov \delta\Xi^a}-\left({\delta a_{2}^{ab}(V)\ov\delta \Xi^c}\Xi^b+a_2^{ab}(\Xi)\delta^{bc}\right){\L}^a{\delta\S\ov\delta\L^c}\Bigg) +c.c.\right] \nonumber \\
		&-&\frac{1}{8}\left({\delta F_{1}^{ab}(V)\ov\delta V^c}V^b+F_1^{ab}(V)\delta^{bc}\right)G^c_V(D^2c'^a+\ovl{D}^2 \ovl{c}'^a)-\frac{1}{8}A^a D^2(F_{1}^{ab}    V^b)-\frac{1}{8}\ovl A^a \ovl D^2(F_{1}^{ab}    V^b).
		\label{ii}
	\end{eqnarray}
	
	\section{Analysis of the counterterm and renormalization factors \label{is}}

	Having determined the most general form of the local invariant counterterm, eq.\eqref{ii}, we observe, however,  that the terms on the last line, 
	
	\beq
	&-&\frac{1}{8}S_{\S}(c'^a D^2 (F_1^{ab}V^b)+\ovl{c}'^a \ovl{D}^2(F_1^{ab}V^b)) \nonumber \\
	&=&-\frac{1}{8}\left({\delta F_{1}^{ab}(V)\ov\delta V^c}V^b+F_1^{ab}(V)\delta^{bc}\right)G^c_V(D^2c'^a+\ovl{D}^2 \ovl{c}'^a)-\frac{1}{8}A^a D^2(F_{1}^{ab}    V^b)-\frac{1}{8}\ovl A^a \ovl D^2(F_{1}^{ab}    V^b), \label{aabbcc}
	\eeq cannot be rewritten in an exact parametric form in terms of the starting action $\Sigma$. This feature is due to dependence of the gauge fixing on the dimensionless field $V$. As a consequence, the renormalization of the gauge fixing itself is determined up to an ambiguity of the type of eq.\eqref{eqq}. As was mentioned before, this term can be handled by starting with the generalized gauge fixing of eq.\eqref{eqq}. This means that we could have equally started with a term in the action like
	\beq
	s\left[\frac{1}{8}tr\,\int dV(c'D^{2}V+\overline{c}'\,\overline{D}^{2}V)\right]\rightarrow s\left[\frac{1}{8}tr\,\int dV(c'D^{2}\mathcal F(V)+\overline{c}'\,\overline{D}^{2}\mathcal F^a(V))\right],  \label{ggfx}
	\eeq
 with $\mathcal{F}^a$ given by eq.\eqref{eqq}. Since $\mathcal{F}^a$ is now a composite field, we need to introduce it into the starting action through a suitable external source. In order to maintain BRST invariance, we make use of a BRST doublet of external sources $(R^a,P^a)$, of dimension 2, R-weight 0 and ghost number $(-1,0)$
	\begin{eqnarray}
		sR^a=P^a, \,\,\, sP^a=0
	\end{eqnarray}
	and introduce the term
	\begin{eqnarray}
		\int dV s(R^a \mathcal{F}^a(V))=\int dV \left(P^a \mathcal{F}^a(V)-R^a\frac{\partial \mathcal{F}^a}{\partial V^c}G^{c}(V)\right),
	\end{eqnarray}
	so that the full action is now given by
	\begin{eqnarray}
		\Sigma_{{\rm SPYM}} & = & -\frac{1}{128g^{2}}tr\int dS\Bigg(W^{a}W_{a}\Bigg)+\frac{m^2}{2}\int dV\, {\cal{M}}(V^H) \nonumber \\
		& + & \frac{1}{8}\,\int dS\Bigg\{ A^{a}\ovl{D}^2 D^{2}\mathcal{F}^{a}(V)-c'^{a}\ovl{D}^2 D^{2}\Bigg[\frac{\partial \mathcal{F}^a}{\partial V^c}G^{c}_{V}\Bigg]+c.c. \Bigg\} \nonumber \\
		&+&\frac{1}{8}\int dS \,\Bigg\{B\ovl{D}^2D^2 V^H -\eta'^a\ovl D^2 D^2 G^{a}_{V^H} +c.c. 
		\Bigg\} \nonumber \\
		&  +&\int dS\Bigg\{-\Lambda^{a}G_{\Xi}^a  +\frac{i}{2}f^{abc}L^{a}c^{b}c^{c}+c.c\Bigg\} \nonumber\\
		&+&\int dV \left(\Omega^{a}G_{V}^a +\Pi^a V^{H,a}+\P^{a}G_{V^H}^a +P^a \mathcal{F}^a(V)-R^a\frac{\partial \mathcal{F}^a(V)}{\partial V^c}G^{c}_{V}\right).  \label{nnacctt}
	\end{eqnarray}
	
	The action $\Sigma_{{\rm SPYM}}$ obeys the following Ward identities:
	\begin{itemize}
		\item The Slavnov-Taylor identity: 
		
		\begin{eqnarray}
			\mathcal{S}(\Sigma)
			=\int dV \left\{ \frac{\delta \Sigma}{\delta \Omega^a}\frac{\delta \Sigma}{\delta V^a}+ P^a\frac{\delta \Sigma}{\delta R^{a}} \right\}
			+ \left( \int dS  \frac{\delta\Sigma}{\delta\Lambda^{a}}\frac{\delta\Sigma}{\delta\Xi^{a}} +\frac{\delta \Sigma}{\delta L^a}\frac{\delta \Sigma}{\delta c^a}+ A^a \frac{\delta \Sigma}{\delta c'^a} + c.c. \right) = 0
		\end{eqnarray}
		
		\item The gauge-fixing equations:
		
		\begin{eqnarray}
			\frac{\delta \Sigma}{\delta A^a} = \frac{1}{8} \bar{D}^2{D}^2\frac{\delta \S}{\delta P^a} , \qquad \frac{\delta \Sigma}{\delta \bar{A}}= \frac{1}{8} D^2 \bar{D}^2 \frac{\delta \S}{\delta P^a}
			 \nonumber 
		\label{gff2}
		\end{eqnarray}
		\item The equation for the Lagrange multiplier $B^a$:
		\beq
		\frac{\delta \Sigma}{\delta B^a} &=& \frac{1}{8} \bar{D}^2{D}^2 \frac{\delta \Sigma}{\delta\Pi^a}, \qquad \frac{\delta \Sigma}{\delta \bar{B}^{a}}= \frac{1}{8} D^2 \bar{D}^2 \frac{\delta \Sigma}{\delta\Pi^a}
		\eeq

		\item The anti-ghost equations:
		
		\beq
		\mathcal{G}_-^a \Sigma =0, \qquad \bar{\mathcal{G}}_-^a \Sigma =0, 
		\label{ag2}
		\eeq
		with 
		\beq
		\mathcal{G}_-^a= \frac{\delta}{\delta c'^a}-\frac{1}{8} \bar{D}^2 D^2 \frac{\delta }{\delta R^a} \quad \text{and} \quad \bar{\mathcal{G}}_-^a= \frac{\delta}{\delta \bar{c}'^a}-\frac{1}{8} D^2 \bar{D}^2  \frac{\delta }{\delta R^a} 
		\eeq
		
		\item The $\eta$ Ward identities
		
		\beq
		\mathcal{F}_-^a \Sigma =0, \qquad {\mathcal{\ovl F}}_{-}^a \Sigma =0 ,
		\eeq
		with 
	\beq
		\mathcal{F}_-^a= \frac{\delta}{\delta \eta'^a}+\frac{1}{8} \bar{D}^2 D^2 \frac{\delta }{\delta \Psi^a} \quad \text{and} \quad {\mathcal{\ovl F}}_{-}^a= \frac{\delta}{\delta \ovl{\eta}'^a}+\frac{1}{8} D^2 \bar{D}^2  \frac{\delta }{\delta \Psi^a} 
		\eeq
		
		%\item The ghost equation is not linear in the fields anymore, so we do not consider it.
		\item The linear symmetries under supersymmetry, translations, $R$-transformations and rigid transformations are expressed by the Ward identities in \eqref{end}.
		
	\end{itemize}
	Since $R$ and $P$ are a pair of BRST doublet, they do not appear in the non-trivial part of the counterterm  \cite{Piguet:1995er}, so this will remain as in equation \eqref{d}. On the other hand, the $\Delta^{(-1)}$-term becomes
	
	\beq
	\Delta^{-1}&=& \int dV \Bigg( F_{1}^{ab}V^b \O^a + F_{2}^{ab}V^b D^2 c'^a +F_{3}^{ab}D^2 V^b c'^a+D^2 F_{4}^{ab} V^b c'^a \nonumber \\
	&&+F_8^{ab}V^b R^a +c.c\Bigg) 
	+\left[ \int dS \left( a_1^{ab} L^a c^b +a_{2}^{ab}\Xi^{a}\Lambda^{b} \right) +c.c.\right].
	\label{hm}
	\eeq
	In a similar fashion as in the analysis of eqs.\eqref{iii}--\eqref{mm}, we can use the relation $[\frac{\partial }{\partial A^a}-\frac{1}{8}\ovl D^2   D^2 \frac{\delta}{\delta P^a},S_{\Sigma} ]=\frac{\delta}{\delta c'^a}-\frac{1}{8}\ovl D^2 D^2 \frac{\delta}{\delta R}$, to find
	
	\beq
	\left(\frac{\delta}{\delta c'^a}-\frac{1}{8}\ovl D^2 D^2 \frac{\delta}{\delta R}\right)\Delta^{-1}&=& \ovl D^2 D^2(F_2^{ab} V^b)+\ovl D^2 (F_3^{ab} D^2 V^b)+\ovl D^2 \left((D^2 F_4^{ab})V^b\right)-\frac{1}{8}\ovl D^2 D^2 (F_8^{ab} V^b)=0,
	\eeq
	from which we obtain the relations 
	\beq
	F_2^{ab}&=&\frac{1}{8}F_8^{ab} \nonumber \\
	F_3^{ab}&=&0 \nonumber \\
	F_4^{ab}&=&a \delta^{ab}, \label{rrllt}
	\eeq  
	so that \eqref{hm} becomes 
	\beq
	\Delta^{-1}&=& \int dV \Bigg( F_{1}^{ab}V^b \O^a + F_{2}^{ab}V^b(R^a+\frac{1}{8} D^2 c'^a  +\frac{1}{8}\ovl D^2 \ovl c'^a)\Bigg) \nonumber \\
	&&+\left[ \int dS \left( a_1 L^a c^a+a_{2}^{ab}\Xi^{b}\Lambda^{a} \right) +c.c.\right].
	\eeq
	We now set
	\beq
	F_{2}^{ab}(V)&=&F_{2}^{ab}(0)+\til {F}^{ab}_{2}(V), \nonumber \\
	\til {F}^{ab}_{2}(V)&=& \til \a_1^{abc}V^c+\til \a_2^{abcd}V^cV^d+\til \a_2^{abcde}V^cV^dV^e+...\, ,
	\eeq
	where $F_2^{ab}(0)=F_2(0)\delta^{ab}$ is the first, $V$-independent term of the Taylor expansion of $F_2^{ab}(V)$ in powers of $V$, and $\til F^{ab}_2 (V)$ denotes the remaining $V$-dependent terms. We find that $F_2^{ab}(0)$ is connected to the renormalization of the fields within the gauge-fixing, while $\til F^{ab}_2(V)$ renormalizes the gauge parameters in eq.\eqref{eqq}. Employing the generalized gauge-fixing \eqref{ggfx}, we can now write the full counterterm in a complete parametric form, namely 
	\beq
	\S_{CT}&=&\int dS a_0 g^2 \frac{\delta \Sigma}{\delta g^2} + b_0\int dV \,m^2 \frac{\delta \S}{\delta m^2}+b_1 \left[ \int dS \left(B^a \frac{\delta\S}{\delta B^a}\right) +c.c.\right] + b_1 \int dV \Pi^a\frac{\delta\S}{\delta \Pi^a} \nonumber \\
	&+&\frac{1}{2}b_1 \int dV \Psi^a \frac{\delta\S}{\delta \Psi^a}+\frac{1}{2}b_1\left[ \int dS \left(\eta^a\frac{\delta\S}{\delta \eta^a}\right) +c.c.\right]+\frac{1}{2}b_1\left[ \int dS \left(\eta'^a\frac{\delta\S}{\delta \eta'^a}\right) +c.c.\right] \nonumber \\
	&+&\int dV \left(\tilde{\s}_1^{abc}\frac{\delta \Sigma}{\delta \sigma_1^{abc}}+\tilde{\s}_2^{abcd}\frac{\delta \Sigma}{\delta \sigma_2^{abcd}}+\tilde{\s}_3^{abcde}\frac{\delta \Sigma}{\delta \sigma^{abcde}_3}+...\right) \nonumber \\
	&-&\int dV\Bigg(\left({\delta F_{1}^{ab}\ov\delta V^c}V^b+F_1^{ab} \delta^{bc}\right) {\O}^a{\delta\S\ov\delta\O^c}+F_{1}^{ab}V^b{\delta\S\ov\delta V^a}+F_2(0)P^a\frac{\delta \S}{\delta P^a}+F_2(0)\ovl A^a \frac{\delta \S}{\delta \ovl A^a} \nonumber \\
	&+&F_2(0) A^a \frac{\delta \S}{\delta  A^a}+F_2(0)R^a\frac{\delta \S}{\delta R^a}+F_2(0)c'^a\frac{\delta \S}{\delta c'^a}+F_2(0)\ovl c'^a\frac{\delta \S}{\delta \ovl c'^a}\Bigg) \nonumber \\
	&+& \left[ \int dS\Bigg( a_1^{ab} L^a\frac{\delta \S}{\delta L^b}-a_1^{ab}c^a\frac{\delta \S}{\delta c^b}-\left({\delta a_{2}^{ab}(\Xi)\ov\delta \Xi^c}\Xi^b+a_2^{ab}(\Xi)\delta^{bc}\right){\L}^a{\delta\S\ov\delta\L^c} + a_2^{ab}(\Xi)\Xi^b{\delta \S\ov \delta\Xi^a}\Bigg) +c.c.\right] \nonumber \\
	&+&\int dV (\til \a_1^{abc}-F_2(0)a_1^{abc})\frac{\delta \S}{\delta \a_1^{abc}}+(\til \a_2^{abcd}-F_2(0)\a_2^{abcd})\frac{\delta \S}{\delta \a_2^{abcd}}+(\til \a_3^{abcde}-F_2(0)\a_3^{abcde})\frac{\delta \S}{\delta \a_3^{abcde}}+...\, ,
	\label{yy1}
	\eeq
	where the dots ... in the last line denote the infinite set of terms of the kind
	\beq
	\sum_j(\til \a_j^{abcde..}-F_2(0)\a_j^{abcde..})\frac{\delta \S}{\d \a_j^{abcde}}, \,\,\,\,\,\, j=4,...,\infty.
	\eeq
	The usefulness of rewriting the counterterm \eqref{yy1} in the parametric form becomes clear by casting it into the form 
	\beq
	\mathcal{R}\S=0,
	\eeq
	with
  \beq
	\mathcal{R}&=&\int dS a_0 g^2 \frac{\delta }{\delta g^2}  + b_0\int dV \,m^2 \frac{\delta }{\delta m^2}+b_1 \left[ \int dS \left(B^a \frac{\delta}{\delta B^a}\right) +c.c.\right] + b_1 \int dV \Pi^a\frac{\delta}{\delta \Pi^a} \nonumber \\
	&+&\frac{1}{2}b_1 \int dV \Psi^a \frac{\delta}{\delta \Psi^a}+\frac{1}{2}b_1\left[\int dS \left(\eta^a\frac{\delta}{\delta \eta^a}\right) +c.c.\right]+\frac{1}{2}b_1\left[\int dS \left(\eta'^a\frac{\delta}{\delta \eta'^a}\right) +c.c.\right] \nonumber \\
	&+&\int dV \left(\tilde{\s}_1^{abc}\frac{\delta }{\delta \sigma_1^{abc}}+\tilde{\s}_2^{abcd}\frac{\delta }{\delta \sigma_2^{abcd}}+\tilde{\s}_3^{abcde}\frac{\delta }{\delta \sigma_3^{abcde}}+...\right) \nonumber \\
	&-&\int dV\Bigg(\left({\delta F_{1}^{ab}\ov\delta V^c}V^b+F_1^{ab} \delta^{bc}\right) {\O}^a{\delta\ov\delta\O^c}+F_{1}^{ab}V^b{\delta\ov\delta V^a}+F_2(0)P^a\frac{\delta }{\delta P^a}+F_2(0)\ovl A^a \frac{\delta }{\delta \ovl A^a}\nonumber \\
	&+&F_2(0) A^a \frac{\delta }{\delta  A^a}+F_2(0)R^a\frac{\delta }{\delta R^a}+F_2(0)c'^a\frac{\delta }{\delta c'^a}+F_2(0)\ovl c'^a\frac{\delta }{\delta \ovl c'^a}\Bigg) \nonumber \\
	&+& \left[\int dS\Bigg( a_1^{ab} L^a\frac{\delta }{\delta L^b}-a_1^{ab}c^a\frac{\delta }{\delta c^b}-\left({\delta a_{2}^{ab}(\Xi)\ov\delta \Xi^c}\Xi^b+a_2^{ab}(\Xi)\delta^{bc}\right){\L}^a{\delta\ov\delta\L^c} + a_2^{ab}(\Xi)\Xi^b{\delta \ov \delta\Xi^a}\Bigg) +c.c.\right]\nonumber \\
	&+&\int dV (\til \a_1^{abc}-F_2(0)\a_1^{abc})\frac{\delta}{\delta \a_1^{abc}}+(\til \a_2^{abcd}-F_2(0)\a_2^{abcd})\frac{\delta }{\delta \a_2^{abcd}}+(\til \a_3^{abcde}-F_2(0)\a_3^{abcde})\frac{\delta }{\delta \a_3^{abcde}}+...\, .
	\label{yy}
	\eeq
	Now, in order to determine the renormalization factors we can use that
	\beq
	\S(\Phi)+\varepsilon \S_{CT}(\Phi)=\S(\Phi)+\varepsilon \mathcal{R}\S(\Phi)=\Sigma(\Phi_0)+\mathcal{O}(\varepsilon^2),
	\eeq
	with
	\beq
	\Phi_0=Z_{\Phi}\Phi=(1+\varepsilon \mathcal{R})\Phi+\mathcal{O}(\varepsilon^2),
	\eeq
	where $\Phi_0$ is a short-hand notation for all renormalized quantities: fields, parameters and external sources.  We thus find the  the following renormalization factors
	\beq
	Z_g=1+a_0\,\nonumber\\
	Z_{m^2}=1+b_0\,\nonumber\\
	Z_B=Z_{\ovl B}=Z_{\Pi}=Z_{\Psi}^2=Z_{\eta}^2=Z_{\eta'}^2=Z_{\ovl \eta}^2=Z_{\ovl \eta'}^2=1+b_1\,\nonumber\\
	Z_{\O}^{ab}=\delta^{ab}-({\delta F_{1}^{ac}\ov\delta V^b}V^c+F_1^{ab})\,\nonumber\\
	Z_{V}^{ab}=\delta^{ab}+F_1^{ab}\,\nonumber\\
	Z_P=Z_A=Z_{\ovl A}=Z_R=Z_{c'}=Z_{\ovl c'}=1+F_2(0)\,\nonumber\\
	Z_L^{ab}=Z_{\ovl L}^{ab}=Z_c^{ab}=Z_{\ovl c}^{ab}=\delta^{ab}+a_1^{ab}\,\nonumber\\
	Z_{\L}^{ab}=\delta^{ab}-({\delta a_{2}^{ac}\ov\delta \Xi^b}\Xi^c+a_2^{ab}(\Xi))\,\nonumber\\
	Z_{\ovl \L}^{ab}=\delta^{ab}-({\delta \ovl a_{2}^{ac}\ov\delta \ovl \Xi^b}\ovl \Xi^c+\ovl a_2^{ab}(\ovl \Xi))\,\nonumber\\
	Z_{\Xi}^{ab}=\delta^{ab}+a_2^{ab}(\Xi)\,\nonumber\\
	Z_{\ovl \Xi}^{ab}=\delta^{ab}+\ovl a_2^{ab}(\ovl \Xi),
	\eeq
	as well as a multiplicative renormalization of the infinite set of gauge parameters $(\s_1^{abc},\s_2^{abcd},\s_3^{abcde})$ and $(\a_1^{abc},\a_2^{abcd},\a_3^{abcde})$ of  equations \eqref{tit} and \eqref{eqq}, being
	\beq
	(\s_1^{abc})_0&=&(1+\varepsilon \til \s_1^{abc})\s_1^{abc} \nonumber \\
		(\s_2^{abcd})_0&=&(1+\varepsilon \til \s_2^{abce})\s_2^{abcd}\nonumber \\
			(\s_3^{abcde})_0&=&(1+\varepsilon \til \s_3^{abcde})\s_3^{abcde} \nonumber \\
			&...&
	\eeq
	and
		\beq
	(\a_1^{abc})_0&=&(1+F_2(0))\a_1^{abc}+\varepsilon \til \a_1^{abc} \nonumber \\
		(\a_2^{abcd})_0&=&(1+F_2(0))\a_2^{abcd}+\varepsilon \til \a_2^{abcd}\nonumber \\
	(\a_3^{abcde})_0&=&(1+F_2(0))\a_3^{abcde}+\varepsilon \til \a_3^{abcde} \nonumber \\
	&...&
	\label{kal}
	\eeq
	This shows that the inclusion of the generalized field $\mathcal{F}(V)$ in the gauge fixing leads to the standard renormalization of the fields, parameters and sources. The renormalization of $\mathcal{F}(V)$ itself is encoded in the renormalization of the infinite set of gauge parameters $(\a_1^{abc},\a_2^{abcd},\a_3^{abcde}, ....)$, as in eq. \eqref{kal}. \\
	Note that both $V$ and $\Xi$, as well as their sources $\O$ and $\L$, are renormalized in a non-linear way through a power series in $V$ and $\Xi$, respectively. This is expected, due to the fact that both superfields are dimensionless. However, one has to note that the dimensionless superfield $V$ contains a massive supermultiplet $(A_{\m},\l)$. Despite the fact that $V$ itself renormalizes in a non-linear way due to its dimensionless nature, the component fields $(A_{\m},\l)$ do renormalize in fact in a standard multiplicative way through a constant ({\it i.e.} field independent) renormalization factors, a feature which can be checked out by employing the the Wess-Zumino gauge.

\section{Conclusion}
In this work we took a first step towards the understanding of Stueckelberg-like models in supersymmetric non-abelian gauge theories. The gauge invariant transverse field configuration $V^H$ has been investigated in supersymmetric Yang-Mills theory with the Landau gauge. An auxiliary chiral superfield $\Xi$ was introduced that compensates the gauge variation of the vector superfield $V$, thus preserving gauge invariance of the composite field $V^H$. This gauge invariant composite field allows  the construction of a local BRST-invariant massive model, summarized by the action \eqref{o1}. Both $V$ and $\Xi$ are dimensionless, which leads to ambiguities in defining both the mass term and the gauge fixing term. However, working with a generalized gauge fixing term, we find that the model turns out to  be renormalizable to all orders of perturbation theory, as was discussed in sections \ref{het} and \ref{is}. 
\\
 As a possible future application of the present result, let us mention that the possibility of having constructed a manifestly BRST invariant supersymmetric renormalizable version of the modified Stueckelberg models introduced in  \cite{Capri:2015ixa,  Capri:2015nzw, Capri:2016aqq, Fiorentini:2016rwx, Capri:2016gut,Capri:2018} can open the possibility to investigate the important issue of the non-perturbative phenomenon of the Gribov copies directly in superspace, by generalizing to ${\cal N}=1$ the Gribov-Zwanziger setup. This would enable us to study aspects of  the non-perturbative region of ${\cal N}=1$ confining supersymmetric theories, see also \cite{Amaral:2013uya} for a preliminary attempt in this direction.  

\section*{Acknowledgements}

The Conselho Nacional de Desenvolvimento Científico e Tecnológico (CNPq-Brazil), the Faperj, Fundação de
Amparo à Pesquisa do Estado do Rio de Janeiro, the SR2-UERJ and the Coordenação de Aperfeiçoamento
de Pessoal de Nível Superior (CAPES) are gratefully acknowledged for financial support. S. P. Sorella is
a level PQ-1 researcher under the program Produtividade em Pesquisa-CNPq, 300698/2009-7; M. A. L. Capri
is a level PQ-2 researcher under the program Produtividade em Pesquisa-CNPq, 307783/2014-6. M. S. Guimaraes is supported by the Jovem
Cientista do Nosso Estado program - FAPERJ E-26/202.844/2015, is a level PQ-2 researcher under the
program Produtividade em Pesquisa-CNPq, 307905/2014-4 and is a Procientista under SR2-UERJ.

	\appendix
	
	\section{Notation}

	\begin{eqnarray}
		V^{H} & = & V^{H\,a}T^{a}\nonumber\\
		\left[T^{a},T^{b}\right] & = & -if^{acb}T^{c}\nonumber\\
		\int dV & = & \int d^{4}xd^{2}\theta d^{2}\bar{\theta}\nonumber\\
		\int dS & = & \int d^{4}xd^{2}\nonumber\theta\\
		\int d\bar{S} & = & \int d^{4}xd^{2}\bar{\theta}
	\end{eqnarray}
	
\section{N=1 Superfields\label{apb}}

In the $\mathcal{N}=1$ case, we have a Poincaré algebra  with spinor charges that anticommutes as
\beq
\{Q_{\a},\ovl{Q}_{\dot{\a}}\}=2\sigma_{\a\dot\a}^{\m}P_{\m},\,\, \{Q_{\a},Q_{\b}\}=0, \,\, \{\ovl Q_{\dot\a},\ovl Q_{\dot\b}\}=0,
\eeq 
and commutes
\beq
[Q_{\a},M_{\m\n}]&=&\ha(\sigma_{\m\n})_{\a}^{\,\,\b}Q_{\b},\,\,[\ovl Q^{\dot\a},M_{\m\n}]=-\ha(\ovl\sigma_{\m\n})_{\,\,\dot\b}^{\dot\a}\ovl Q^{\dot \b},
\eeq
\beq
[Q_{\a},P_{\m}]&=& 0,\,\, [\ovl Q_{\dot{\a}},P_{\m}]=0,
\eeq
and to
\beq
[Q_{\a} , R]=-Q_{\a},\,\, [\ovl Q_{\dot{\a}} , R]=\ovl Q_{\dot \a}
\eeq
The $R$ is a symmetry transforming different charges in a theory into each other and that is isomorphic to a global $U(1)$ group.   

Using the supersymmetry transformations as a Lie algebra we can define objects 
\beq
G(x,\theta,\ovl \theta)=e^{i(-x^\mu P_{\mu}+\theta Q + \ovl \theta \ovl Q)}
\eeq
called superfields which transform covariantly under supersymmetry transformations. Superfields can be of the general type, or chiral type. A chiral superfield $A(x,\theta,\ovl \theta)$ and anti-chiral superfield $A(x,theta,\ovl \theta)$, is a superfield obeying the constraint
\beq
\ovl D_{\dot \a}A=0, \,\,D_{\a}\ovl A=0,
\eeq
where 
\beq
D_{\a}=\frac{\d}{\d\t^\a}-i\s_{\a\dot\a}^{\m}\ovl\t^{\dot\a}\d_{\m},\,\,\ovl D_{\dot\a}=-\frac{\d}{\d\ovl\t^{\dot\a}}+i\t^\a\s_{\a\dot\a}^{\m}\d_{\m}
\eeq
are the covariant superspace derivatives. The superfields are functions of superspace which should be understood in components by series power in $\t$ and $\ovl \t$. The transformation laws (translation, supersymmetry and R-symmetry) of a superfield $\f$ are respectively defined 
\beq
\delta_{\m}^{P}\f &=& \partial_{\m}\f ,\\
\delta_{\a}^{Q}\f &=& \left(\frac{\d}{\d\t^\a}+i\s_{\a\dot\a}^{\m}\ovl\t^{\dot\a}\d_{\m}\right)\f,\\
\delta_{\dot\a}^{\ovl Q}\f &=&\left(-\frac{\d}{\d\ovl\t^{\dot\a}}-i\t^\a\s_{\a\dot\a}^{\m}\d_{\m}\right)\f,
\eeq
and
\beq
\delta^{R}\f = i\left(n+\t^\a\frac{\d}{\d\t^\a}-\ovl\t^{\dot\a}\frac{\d}{\d\ovl\t^{\dot\a}}\right)\f .\label{b12}
\eeq
The $n$ number in \eqref{b12} is the ``R-weight'' of the superfield $\f$. The R-weights are opposite to each other in complex conjugates superfields. These operators obey the super-Poincaré algebra
\beq
\{\delta^{Q}_{\a},\delta^{\ovl{Q}}_{\dot{\a}}\}&=&-2i\sigma_{\a\dot\a}^{\m}\delta^{P}_{\m},\\
\  [\delta_{\a}^{Q},\delta^{R}] &=&i\delta_{\a}^{Q},\,\,[\delta_{\dot\a}^{\ovl Q},\delta^{R}]=-i\delta_{\dot\a}^{\ovl Q}.
\eeq

\section{$\mathcal{N}=1$ Supersymmetric Yang-Mills\label{apc}} 
For the benefit of the reader, we provide in this appendix a short  overview of the well known renormalizability of pure ${\cal N}=1$ standard massless Super Yang-Mills in the Landau gauge. Let us start by giving the complete BRST invariant action, namely 
\begin{eqnarray}
{\S}_{{\rm SYM}} & =& \S_{SYM}+\S_{SGF}+\S_{EXT}  \nonumber \\
& = &  -\frac{1}{128g^{2}}tr\int dS \,\,\left(W^{a}W_{a}\right)+s\Bigg[\frac{1}{8}\,\int dV(c'^a D^{2}V^a+\overline{c}'^a\,\overline{D}^{2}V^a)\Bigg] \nonumber \\
& + & \int dV \Omega^{a}G^a_V(c, \ovl c) +\frac{i}{2}\int dSf^{abc}L^{a}c^{b}c^{c} 
+\frac{i}{2}\int d\ovl S f^{abc}\bar{L}^{a}\bar{c}^{b}\bar{c}^{c} \;. \label{aaaccctttt}
\end{eqnarray}
The full action $\Sigma_{SPYM}$ obeys the Ward identities \eqref{gff},\eqref{ag},\eqref{gh},\eqref{end} as well as the Slavnov-Taylor identity: 

\begin{eqnarray}
\mathcal{S}(\Sigma)
=\int dV \left\{ \frac{\delta \Sigma}{\delta \Omega^a}\frac{\delta \Sigma}{\delta V^a}\right\} 
+  \left[\int dS \left(  \frac{\delta \Sigma}{\delta L^a}\frac{\delta \Sigma}{\delta c^a}+ A^a \frac{\delta \Sigma}{\delta c'^a} \right) +c.c.\right]= 0 \;. 
\end{eqnarray}
As usual, the counterterm $\S_{CT}$ can be written as  
\begin{equation} 
\S_{CT}=\Delta+\mathcal{B}_{\S}\Delta^{(-1)}   \;, \label{ct}
\end{equation}
with 
\beq
\Delta=a_0 \,\, Tr\int dS \,\, W^{\a}W_{\a},
\eeq
and
\beq
\Delta^{-1}&=& \int dV \left( F_{1}^{ab}(V)V^b \O^a -{1\ov 8} F_{1}^{ab}(V)V^b(D^2 c'^a +\ovl D^2 \ovl c'^a)\right) 
+\left[\int dS \, \left( a_1^{ab} L^a c^b \right) +c.c.\right].
\eeq
When constructing the counterterm, we will run into to same parametrizing problem as in section \ref{het}. Repeating the procedure of section \ref{is}, we introduce a doublet $(R^a,P^a)$ and the full action is now given by
\begin{eqnarray}
\Sigma_{{\rm SPYM}} & = & -\frac{1}{128g^{2}}tr\int dS\left(W^{a}W_{a}\right)+ \left[\int dS\Bigg(\frac{i}{2}f^{abc}L^{a}c^{b}c^{c}\Bigg) +c.c.\right] \\
& + & \frac{1}{8}\,\int dV\Bigg\{ A^{a}D^{2}\mathcal{F}^{a}-c'^{a}D^{2}\Bigg[\frac{\partial \mathcal{F}^a}{\partial V^c}G^{c}_{V}\Bigg]+c.c. \Bigg\} \nonumber \\
&+&\int dV \left(\O^a G^{a} _{V}( c, \ovl c) +P^a \mathcal{F}^a(V)-R^a\frac{\partial \mathcal{F}^a}{\partial V^c}G^{c}_{V}(c, \ovl c)\right).
\end{eqnarray}
The action obeys the Ward identities \eqref{gff2},\eqref{ag2},\eqref{end2} as well as the Slavnov-Taylor identity:
\begin{eqnarray}
\mathcal{S}(\Sigma)
=\int dV \left\{ \frac{\delta \Sigma}{\delta \Omega^a}\frac{\delta \Sigma}{\delta V^a}+ P^a\frac{\delta \Sigma}{\delta R^{a}} \right\} 
+ \left[ \int dS  \left( \frac{\delta \Sigma}{\delta L^a}\frac{\delta \Sigma}{\delta c^a}+ A^a \frac{\delta \Sigma}{\delta c'^a} \right) +c.c.\right]  = 0   \;. \label{nstnst}
\end{eqnarray}
Making use of the relation $[\frac{\partial }{\partial A^a}-\frac{1}{8}\ovl D^2   D^2 \frac{\delta}{\delta P^a},S_{\Sigma} ]=\frac{\delta}{\delta c'^a}-\frac{1}{8}\ovl D^2 D^2 \frac{\delta}{\delta R}$, we find
\beq
\Delta^{-1}&=& \int dV \Bigg( F_{1}^{ab}V^b \O^a + F_{2}^{ab}V^b(R^a+\frac{1}{8} D^2 c'^a  +\frac{1}{8}\ovl D^2 \ovl c'^a)\Bigg) +\left[ \int dS \, \left( a_1^{ab} L^a c^b \right) +c.c.\right].
\eeq
Defining 
\beq
F_2^{ab}(V)&=&F_2(0)\delta^{ab}+\til F_2^{ab}(V) \nonumber \\
\til F_2^{ab}(V)&=& \til \a_1^{abc}V^c+\til \a_2^{abcd}V^d+\til \a_3^{abcde}V^e+...\, ,
\eeq
we find the counterterm to be 
\beq
\S_{CT}&=&\int dS a_0 g^2 \frac{\delta \Sigma}{\d g^2}+\int dV\Bigg(F_2(0)P^a\frac{\delta \S}{\delta P^a}+F_2(0)\ovl A^a \frac{\delta \S}{\delta \ovl A^a} \nonumber \\
&+&F_2(0) A^a \frac{\delta \S}{\delta  A^a}+F_2(0)R^a\frac{\delta \S}{\delta R^a}+F_2(0)c'^a\frac{\delta \S}{\delta c'^a}+F_2(0)\ovl c'^a\frac{\delta \S}{\delta \ovl c'^a}\Bigg) \nonumber \\
&+&\int dV \Bigg( F_{1}^{ab}V^b{\delta\S\ov\delta V^a}-\left({\delta  F_{1}^{ab}\ov\delta V^c}V^b+ F_1^{ac}\right) {\O}^a {\delta\S\ov\delta\O^c} \Bigg) \nonumber \\
&+&\left[ \int dS\Bigg( \a_1 L^a\frac{\d \S}{\d L^a}-\a_1c^a\frac{\d \S}{\d c^a}\Bigg) +c.c.\right] \nonumber \\
&+&\int dV \Bigg((\til \a_1^{abc}-F_2(0))\frac{\delta \S}{\d \a_1^{abc}}+(\til \a_2^{abcd}-F_2(0))\frac{\delta \S}{\d \a_2^{abcd}}+(\til \a_3^{abcde}-F_2(0))\frac{\delta \S}{\d \a_3^{abcde}}+...\Bigg)\,.
\eeq
Following the analysis at the end of section \ref{is}, for the renormalization factors we obtain 
\beq
Z_g=1+a_0 \nonumber \\
Z_{\O}^{ab}=\delta^{ab}-({\delta F_{1}^{ac}\ov\delta V^b}V^c+F_1^{ab}) \nonumber \\
Z_{V}^{ab}=\delta^{ab}+F_1^{ab} \nonumber \\
Z_P=Z_A=Z_{\ovl A}=Z_R=Z_{c'}=Z_{\ovl c'}=1+F_2(0) \nonumber \\
Z_L^{ab}=Z_{\ovl L}^{ab}=Z_c^{ab}=Z_{\ovl c}^{ab}=\delta^{ab}+a_1^{ab} \;. 
\eeq
Thereby concluding the proof of the renormalizability of ${\cal N}=1$ pure massless Super Yang-Mills action.


\newpage
	
	\begin{thebibliography}{widestlabel}
		
		%\cite{Ruegg:2003ps}
		\bibitem{Ruegg:2003ps} 
		H.~Ruegg and M.~Ruiz-Altaba,
		``The Stueckelberg field,''
		Int.\ J.\ Mod.\ Phys.\ A {\bf 19}, 3265 (2004)
		doi:10.1142/S0217751X04019755
		[hep-th/0304245].
		%%CITATION = doi:10.1142/S0217751X04019755;%%
		%217 citations counted in INSPIRE as of 13 Nov 2017
		
		
		%\cite{Delbourgo:1987np}
		\bibitem{Delbourgo:1987np} 
		R.~Delbourgo, S.~Twisk and G.~Thompson,
		``Massive Yang-mills Theory: Renormalizability Versus Unitarity,''
		Int.\ J.\ Mod.\ Phys.\ A {\bf 3}, 435 (1988).
		doi:10.1142/S0217751X88000163
		%%CITATION = doi:10.1142/S0217751X88000163;%%
		%68 citations counted in INSPIRE as of 13 Nov 2017  
		
		%\cite{Lowenstein:1972pr}
		\bibitem{Lowenstein:1972pr} 
		J.~H.~Lowenstein and B.~Schroer,
		``Gauge invariance and Ward identities in a massive vector meson model,''
		Phys.\ Rev.\ D {\bf 6}, 1553 (1972).
		doi:10.1103/PhysRevD.6.1553
		%%CITATION = doi:10.1103/PhysRevD.6.1553;%%
		%94 citations counted in INSPIRE as of 13 Nov 2017 
		
		
		
		
		%\cite{Kunimasa:1967zza}
		\bibitem{Kunimasa:1967zza} 
		T.~Kunimasa and T.~Goto,
		``Generalization of the Stueckelberg Formalism to the Massive Yang-Mills Field,''
		Prog.\ Theor.\ Phys.\  {\bf 37}, 452 (1967).
		doi:10.1143/PTP.37.452
		%%CITATION = doi:10.1143/PTP.37.452;%%
		%93 citations counted in INSPIRE as of 13 Nov 2017
		
		
		%\cite{Slavnov:1970tk}
		\bibitem{Slavnov:1970tk} 
		A.~A.~Slavnov and L.~D.~Faddeev,
		``Massless and massive yang-mills field. (in russian),''
		Theor.\ Math.\ Phys.\  {\bf 3}, 312 (1970)
		[Teor.\ Mat.\ Fiz.\  {\bf 3}, 18 (1970)].
		doi:10.1007/BF01031585
		%%CITATION = doi:10.1007/BF01031585;%%
		%65 citations counted in INSPIRE as of 13 Nov 2017
		
		
		%\cite{Slavnov:1972qb}
		\bibitem{Slavnov:1972qb} 
		A.~A.~Slavnov,
		``Massive gauge fields,''
		Teor.\ Mat.\ Fiz.\  {\bf 10}, 305 (1972).
		%%CITATION = TMFZA,10,305;%%
		%28 citations counted in INSPIRE as of 13 Nov 2017     
		
		
		
		%\cite{Delbourgo:1975uf}
		\bibitem{Delbourgo:1975uf} 
		R.~Delbourgo,
		``A Supersymmetric Stuckelberg Formalism,''
		J.\ Phys.\ G {\bf 1}, 800 (1975).
		doi:10.1088/0305-4616/1/8/003
		%%CITATION = doi:10.1088/0305-4616/1/8/003;%%
		%7 citations counted in INSPIRE as of 14 Nov 2017   
		
		
		%\cite{Kors:2004dx}
		\bibitem{Kors:2004dx} 
		B.~Kors and P.~Nath,
		``A Stueckelberg extension of the standard model,''
		Phys.\ Lett.\ B {\bf 586}, 366 (2004)
		doi:10.1016/j.physletb.2004.02.051
		[hep-ph/0402047].
		%%CITATION = doi:10.1016/j.physletb.2004.02.051;%%
		%161 citations counted in INSPIRE as of 14 Nov 2017
		
		
		%\cite{Kors:2004ri}
		\bibitem{Kors:2004ri} 
		B.~Kors and P.~Nath,
		``A Supersymmetric Stueckelberg U(1) extension of the MSSM,''
		JHEP {\bf 0412}, 005 (2004)
		doi:10.1088/1126-6708/2004/12/005
		[hep-ph/0406167].
		%%CITATION = doi:10.1088/1126-6708/2004/12/005;%%
		%101 citations counted in INSPIRE as of 14 Nov 2017
		
		
		%\cite{Kors:2005uz}
		\bibitem{Kors:2005uz} 
		B.~Kors and P.~Nath,
		``Aspects of the Stueckelberg extension,''
		JHEP {\bf 0507}, 069 (2005)
		doi:10.1088/1126-6708/2005/07/069
		[hep-ph/0503208].
		%%CITATION = doi:10.1088/1126-6708/2005/07/069;%%
		%124 citations counted in INSPIRE as of 14 Nov 2017   
		
		
		%\cite{Nishino:2013kxa}
		\bibitem{Nishino:2013kxa} 
		H.~Nishino and S.~Rajpoot,
		``Variant N= 1 Supersymmetric Non-Abelian Proca-Stueckelberg Formalism in Four Dimensions,''
		Nucl.\ Phys.\ B {\bf 872}, 213 (2013)
		doi:10.1016/j.nuclphysb.2013.03.012
		[arXiv:1304.3482 [hep-th]].
		%%CITATION = doi:10.1016/j.nuclphysb.2013.03.012;%%
		%5 citations counted in INSPIRE as of 14 Nov 2017
		
		
		%\cite{Nishino:2013oea}
		\bibitem{Nishino:2013oea} 
		H.~Nishino and S.~Rajpoot,
		``$N = 1$ supersymmetric ProcaStueckelberg mechanism for extra vector multiplet,''
		Nucl.\ Phys.\ B {\bf 887}, 265 (2014)
		doi:10.1016/j.nuclphysb.2014.08.003
		[arXiv:1309.6393 [hep-th]].
		%%CITATION = doi:10.1016/j.nuclphysb.2014.08.003;%%
		%4 citations counted in INSPIRE as of 14 Nov 2017
		
		
		%\cite{Nishino:2016xsw}
		\bibitem{Nishino:2016xsw} 
		H.~Nishino and S.~Rajpoot,
		``Supersymmetric composite gauge fields with compensators,''
		Phys.\ Lett.\ B {\bf 757}, 237 (2016).
		doi:10.1016/j.physletb.2016.03.062
		%%CITATION = doi:10.1016/j.physletb.2016.03.062;%%    
		
		
		%\cite{Sezgin:2012ka}
		\bibitem{Sezgin:2012ka} 
		E.~Sezgin and L.~Wulff,
		``Supersymmetric Proca-Yang-Mills System,''
		JHEP {\bf 1303}, 023 (2013)
		doi:10.1007/JHEP03(2013)023
		[arXiv:1212.3025 [hep-th]].
		%%CITATION = doi:10.1007/JHEP03(2013)023;%%
		%3 citations counted in INSPIRE as of 14 Nov 2017
		
		
		
		%\cite{Piguet:1996ys}
		\bibitem{Piguet:1996ys} 
		O.~Piguet,
		``Supersymmetry, supercurrent and scale invariance,''
		hep-th/9611003.
		%%CITATION = HEP-TH/9611003;%%
		%21 citations counted in INSPIRE as of 13 Nov 2017
		
		
		%\cite{Piguet:1995zz}
		\bibitem{Piguet:1995zz} 
		O.~Piguet and S.~P.~Sorella,
		``The Antighost equation in N=1 superYang-Mills theories,''
		Phys.\ Lett.\ B {\bf 371}, 238 (1996)
		doi:10.1016/0370-2693(95)01605-8
		[hep-th/9510089].
		%%CITATION = doi:10.1016/0370-2693(95)01605-8;%%
		%6 citations counted in INSPIRE as of 13 Nov 2017
		
		\bibitem{Piguet:1982}
		O.~Piguet and K.~Sibold,``Renormalization of $\mathcal{N}=1$ supersymmetric Yang-Mills theories:(II). The radiative corrections,''
		Nucl. \ Phys. \ B {\bf 197}, 257 (1982)
		doi:10.1016/0550-3213(82)90292-9
		
		
		%\cite{Cucchieri:2004mf}
		\bibitem{Cucchieri:2004mf} 
		A.~Cucchieri, T.~Mendes and A.~R.~Taurines,
		``Positivity violation for the lattice Landau gluon propagator,''
		Phys.\ Rev.\ D {\bf 71}, 051902 (2005)
		doi:10.1103/PhysRevD.71.051902
		[hep-lat/0406020].
		%%CITATION = doi:10.1103/PhysRevD.71.051902;%%
		%93 citations counted in INSPIRE as of 21 Nov 2017
		
		
		%\cite{Cucchieri:2007md}
		\bibitem{Cucchieri:2007md}
		A.~Cucchieri and T.~Mendes,
		``What's up with IR gluon and ghost propagators in Landau gauge? A puzzling
		answer from huge lattices,''
		PoS {\bf LAT2007}, 297 (2007)
		[arXiv:0710.0412 [hep-lat]].
		%%CITATION = POSCI,LAT2007,297;%%
		
		
		%\cite{Cucchieri:2007rg}
		\bibitem{Cucchieri:2007rg}
		A.~Cucchieri and T.~Mendes,
		``Constraints on the IR behavior of the gluon propagator in Yang-Mills
		theories,''
		Phys.\ Rev.\ Lett.\  {\bf 100}, 241601 (2008)
		[arXiv:0712.3517 [hep-lat]].
		%%CITATION = PRLTA,100,241601;%%
		
		%\cite{Cucchieri:2008fc}
		\bibitem{Cucchieri:2008fc}
		A.~Cucchieri and T.~Mendes,
		``Constraints on the IR behavior of the ghost propagator in Yang-Mills
		theories,''
		Phys.\ Rev.\  D {\bf 78}, 094503 (2008)
		[arXiv:0804.2371 [hep-lat]].
		%%CITATION = PHRVA,D78,094503;%%
		
		
		%\cite{Cucchieri:2008mv}
		\bibitem{Cucchieri:2008mv}
		A.~Cucchieri and T.~Mendes,
		``Infrared behavior and infinite-volume limit of gluon and ghost propagators
		in Yang-Mills theories,''
		PoS C {\bf ONFINEMENT8}, 040 (2008)
		[arXiv:0812.3261 [hep-lat]].
		%%CITATION = POSCI,CONFINEMENT8,040;%%
		
		
		
		%\cite{Dudal:2013yva}
		\bibitem{Dudal:2013yva} 
		D.~Dudal, O.~Oliveira and P.~J.~Silva,
		``Källén-Lehmann spectroscopy for (un)physical degrees of freedom,''
		Phys.\ Rev.\ D {\bf 89}, no. 1, 014010 (2014)
		doi:10.1103/PhysRevD.89.014010
		[arXiv:1310.4069 [hep-lat]].
		%%CITATION = doi:10.1103/PhysRevD.89.014010;%%
		%39 citations counted in INSPIRE as of 21 Nov 2017  
		
		
		%\cite{Cornwall:2013zra}
		\bibitem{Cornwall:2013zra} 
		J.~M.~Cornwall,
		``Positivity violations in QCD,''
		Mod.\ Phys.\ Lett.\ A {\bf 28}, 1330035 (2013)
		doi:10.1142/S0217732313300358
		[arXiv:1310.7897 [hep-ph]].
		%%CITATION = doi:10.1142/S0217732313300358;%%
		%23 citations counted in INSPIRE as of 21 Nov 2017 
		
		
		%\cite{Alkofer:2003jj}
		\bibitem{Alkofer:2003jj} 
		R.~Alkofer, W.~Detmold, C.~S.~Fischer and P.~Maris,
		``Analytic properties of the Landau gauge gluon and quark propagators,''
		Phys.\ Rev.\ D {\bf 70}, 014014 (2004)
		doi:10.1103/PhysRevD.70.014014
		[hep-ph/0309077].
		%%CITATION = doi:10.1103/PhysRevD.70.014014;%%
		%243 citations counted in INSPIRE as of 21 Nov 2017
		
		
		%\cite{Dudal:2007cw}
		\bibitem{Dudal:2007cw}
		D.~Dudal, S.~P.~Sorella, N.~Vandersickel and H.~Verschelde,
		``New features of the gluon and ghost propagator in the infrared region from
		the Gribov-Zwanziger approach,''
		Phys.\ Rev.\  D {\bf 77}, 071501 (2008)
		[arXiv:0711.4496 [hep-th]].
		%%CITATION = PHRVA,D77,071501;%%
		
		%\cite{Dudal:2008sp}
		\bibitem{Dudal:2008sp}
		D.~Dudal, J.~A.~Gracey, S.~P.~Sorella, N.~Vandersickel and H.~Verschelde,
		``A refinement of the Gribov-Zwanziger approach in the Landau gauge: infrared
		propagators in harmony with the lattice results,''
		Phys.\ Rev.\  D {\bf 78}, 065047 (2008)
		[arXiv:0806.4348 [hep-th]].
		%%CITATION = PHRVA,D78,065047;%% 
		
		
		
		%\cite{Tissier:2010ts}
		\bibitem{Tissier:2010ts} 
		M.~Tissier and N.~Wschebor,
		``Infrared propagators of Yang-Mills theory from perturbation theory,''
		Phys.\ Rev.\ D {\bf 82}, 101701 (2010)
		doi:10.1103/PhysRevD.82.101701
		[arXiv:1004.1607 [hep-ph]].
		%%CITATION = doi:10.1103/PhysRevD.82.101701;%%
		%91 citations counted in INSPIRE as of 21 Nov 2017
		
		
		%\cite{Strauss:2012dg}
		\bibitem{Strauss:2012dg} 
		S.~Strauss, C.~S.~Fischer and C.~Kellermann,
		``Analytic structure of the Landau gauge gluon propagator,''
		Phys.\ Rev.\ Lett.\  {\bf 109}, 252001 (2012)
		doi:10.1103/PhysRevLett.109.252001
		[arXiv:1208.6239 [hep-ph]].
		%%CITATION = doi:10.1103/PhysRevLett.109.252001;%%
		%93 citations counted in INSPIRE as of 21 Nov 2017 
		
		
		
		%\cite{Capri:2015ixa}
		\bibitem{Capri:2015ixa} 
		M.~A.~L.~Capri {\it et al.},
		``Exact nilpotent nonperturbative BRST symmetry for the Gribov-Zwanziger action in the linear covariant gauge,''
		Phys.\ Rev.\ D {\bf 92}, no. 4, 045039 (2015)
		doi:10.1103/PhysRevD.92.045039
		[arXiv:1506.06995 [hep-th]].
		%%CITATION = doi:10.1103/PhysRevD.92.045039;%%
		%31 citations counted in INSPIRE as of 09 Aug 2017
		
		
		
		
		%\cite{Capri:2015nzw}
		\bibitem{Capri:2015nzw} 
		M.~A.~L.~Capri {\it et al.},
		``More on the nonperturbative Gribov-Zwanziger quantization of linear covariant gauges,''
		Phys.\ Rev.\ D {\bf 93}, no. 6, 065019 (2016)
		doi:10.1103/PhysRevD.93.065019
		[arXiv:1512.05833 [hep-th]].
		%%CITATION = doi:10.1103/PhysRevD.93.065019;%%
		%16 citations counted in INSPIRE as of 09 Aug 2017
		
		
		
		
		
		%\cite{Capri:2016aqq}
		\bibitem{Capri:2016aqq} 
		M.~A.~L.~Capri {\it et al.},
		``Local and BRST-invariant Yang-Mills theory within the Gribov horizon,''
		Phys.\ Rev.\ D {\bf 94}, no. 2, 025035 (2016)
		doi:10.1103/PhysRevD.94.025035
		[arXiv:1605.02610 [hep-th]].
		%%CITATION = doi:10.1103/PhysRevD.94.025035;%%
		%19 citations counted in INSPIRE as of 09 Aug 2017
		
		
		
		%\cite{Fiorentini:2016rwx}
		\bibitem{Fiorentini:2016rwx} 
		M.~A.~L.~Capri, D.~Fiorentini, M.~S.~Guimaraes, B.~W.~Mintz, L.~F.~Palhares and S.~P.~Sorella,
		``Local and renormalizable framework for the gauge-invariant operator $A^2_{\min}$ in Euclidean Yang-Mills theories in linear covariant gauges,''
		Phys.\ Rev.\ D {\bf 94}, no. 6, 065009 (2016)
		doi:10.1103/PhysRevD.94.065009
		[arXiv:1606.06601 [hep-th]].
		%%CITATION = doi:10.1103/PhysRevD.94.065009;%%
		%7 citations counted in INSPIRE as of 09 Aug 2017
		
		
		
		
		
		%\cite{Capri:2016gut}
		\bibitem{Capri:2016gut} 
		M.~A.~L.~Capri, D.~Dudal, A.~D.~Pereira, D.~Fiorentini, M.~S.~Guimaraes, B.~W.~Mintz, L.~F.~Palhares and S.~P.~Sorella,
		``Nonperturbative aspects of Euclidean Yang-Mills theories in linear covariant gauges: Nielsen identities and a BRST-invariant two-point correlation function,''
		Phys.\ Rev.\ D {\bf 95}, no. 4, 045011 (2017)
		doi:10.1103/PhysRevD.95.045011
		[arXiv:1611.10077 [hep-th]].
		%%CITATION = doi:10.1103/PhysRevD.95.045011;%%
		%5 citations counted in INSPIRE as of 13 Nov 2017    
		
		
		%\cite{Gribov:1977wm}
		\bibitem{Gribov:1977wm} 
		V.~N.~Gribov,
		``Quantization of Nonabelian Gauge Theories,''
		Nucl.\ Phys.\ B {\bf 139}, 1 (1978).
		doi:10.1016/0550-3213(78)90175-X
		%%CITATION = doi:10.1016/0550-3213(78)90175-X;%%
		%1563 citations counted in INSPIRE as of 21 Nov 2017
		
		%\cite{Zwanziger:1988jt}
		\bibitem{Zwanziger:1988jt} 
		D.~Zwanziger,
		``Action From the Gribov Horizon,''
		Nucl.\ Phys.\ B {\bf 321}, 591 (1989).
		doi:10.1016/0550-3213(89)90263-0
		%%CITATION = doi:10.1016/0550-3213(89)90263-0;%%
		%103 citations counted in INSPIRE as of 21 Nov 2017    
		
		
		%\cite{Vandersickel:2012tz}
		\bibitem{Vandersickel:2012tz} 
		N.~Vandersickel and D.~Zwanziger,
		``The Gribov problem and QCD dynamics,''
		Phys.\ Rept.\  {\bf 520}, 175 (2012)
		doi:10.1016/j.physrep.2012.07.003
		[arXiv:1202.1491 [hep-th]].
		%%CITATION = doi:10.1016/j.physrep.2012.07.003;%%
		%138 citations counted in INSPIRE as of 21 Nov 2017 
		
		
		%\cite{Piguet:1995er}
		\bibitem{Piguet:1995er}
		O.~Piguet and S.~P.~Sorella,
		%``Algebraic renormalization: Perturbative renormalization, symmetries and anomalies,''
		Lect.\ Notes Phys.\ M {\bf 28} (1995) 1.
		%%CITATION = LNPHA,M28,1;%%
		%143 citations counted in INSPIRE as of 09 Oct 2017

\bibitem{Grisaru:1983}
		S.T.~Gates, M.T.~Grisaru, M.~Rocek (1983) [arXiv: 0108200 [hep-th]]  		
		
		
		\bibitem{Piguet1982}
		O.~Piguet and K.~Sibold,
        Nucl. \ Phys. \ B197 (1982) 257. 
		
		\bibitem{Piguet1984}
		O.~Piguet and K.~Sibold,
        Nucl. \ Phys. \ B247 (1984) 484.
		%\cite{Grisaru:1983}
		
				
		%\cite{Blasi:1989}
		\bibitem{Blasi:1989}
		A. ~Blasi, F. ~Delduc and S. ~P. ~Sorella, 
		Nucl. \ Phys. B {\bf 314}, 409 (1989). doi:10.1016/0550-
		3213(89)90159-4
		
		%\cite{Becchi:1989}
		\bibitem{Becchi:1989}
		C. ~Becchi and O. ~Piguet, Nucl. Phys. B {\bf 315}, 153 (1989). doi:10.1016/0550-
		3213(89)90452-5
		
		%\cite{Capri:2018}
		\bibitem{Capri:2018}
		M. A. L. ~Capri, D.~M. ~van ~Egmond, G. ~Peruzzo, M.~S. ~Guimaraes, O. ~Holanda, S. ~P. ~Sorella, ~R. ~C. ~Terin, ~H. ~C. ~Toledo (2017) [arXiv:1712.04073 [hep-th]]
		
		
		%\cite{Amaral:2013uya}
\bibitem{Amaral:2013uya} 
  M.~M.~Amaral, Y.~E.~Chifarelli and V.~E.~R.~Lemes,
  %``N = 1 Gribov superfield extension,''
  J.\ Phys.\ A {\bf 47}, no. 7, 075401 (2014)
  doi:10.1088/1751-8113/47/7/075401
  [arXiv:1310.8250 [hep-th]].
  %%CITATION = doi:10.1088/1751-8113/47/7/075401;%%
  %4 citations counted in INSPIRE as of 04 Mar 2018
		
		
	\end{thebibliography}
\end{document}